\documentclass[10pt,fleqn]{article}

\usepackage{amsmath,amssymb}
\usepackage{graphicx}
\usepackage[top=0.75in, bottom=0.75in, left=0.75in, right=0.75in, dvips]{geometry}

\newcommand{\unit}[1]{\ensuremath{\text{#1}}}
\newcommand{\dif}[1]{\ensuremath{ \mathrm{d}#1 }}
\newcommand{\borel}{\ensuremath{\mathcal{B}}}
\newcommand{\il}{\ensuremath{\mathcal{L}^{-1}}}
\renewcommand{\Im}{\ensuremath{\mathrm{Im}}}
\newcommand{\atan}{\ensuremath{\mathrm{Arctan}}}
\newcommand{\e}{\ensuremath{\mathrm{e}}}
\newcommand{\ppi}{\ensuremath{\mathrm{\pi}}}
\newcommand{\ii}{\ensuremath{\mathrm{i}}}
\newcommand{\qcd}{\ensuremath{\mathrm{QCD}}}
\newcommand{\erf}{\ensuremath{\mathrm{erf}}}
\def\tref#1#2#3{{#1}~(#2)~#3}

\begin{document}
\title{{\sf A Gaussian Sum-Rules Analysis of Scalar Glueballs}}
\author{
  D.~Harnett,~T.G.~Steele\thanks{email: Tom.Steele@usask.ca}\\
  \textsl{Department of Physics and Engineering Physics}\\
  \textsl{University of Saskatchewan}\\
  \textsl{Saskatoon, Saskatchewan, S7N 5E2}\\
  \textsl{Canada}
 }
\maketitle
\begin{abstract}
Although marginally more complicated than the traditional Laplace sum-rules,
Gaussian sum-rules have the advantage of being able to probe excited and ground states with 
similar sensitivity.
Gaussian sum-rule analysis techniques are applied to the
problematic scalar glueball channel to determine masses, widths and relative resonance strengths
of low-lying scalar glueball states  contributing to the hadronic spectral function.
A feature of our analysis is the inclusion of instanton contributions to the
scalar gluonic correlation function.
Compared with the next-to-leading Gaussian sum-rule, the analysis of the lowest-weighted sum-rule
(which contains a large scale-independent contribution from the low energy theorem)
is shown to be  unreliable because of  instability under QCD uncertainties.
However, the presence of instanton effects leads to approximately consistent
mass scales in the lowest weighted and next-lowest weighted sum-rules.
The analysis of the next-to-leading sum-rule demonstrates that a single narrow 
resonance model does {\em not} provide an adequate description of the hadronic spectral function.  
Consequently, we consider a wide variety of phenomenological models which distribute 
resonance strength over a broad region---some of which lead to excellent agreement 
between the theoretical prediction and phenomenological models. 
Including QCD uncertainties,
our results indicate that the hadronic contributions to the spectral function
stem from a pair of resonances with masses in the range 0.8--1.6 GeV, with the 
lighter of the two potentially having a large width. 
\end{abstract}

\section{Introduction}\label{intro}
Mass predictions for scalar~$(0^{++})$ glueballs extracted from QCD sum-rules
have been problematic mainly due to discrepancies between analyses which are sensitive to the
low-energy theorem for gluonic correlation functions and those which are 
insensitive to this 
quantity~\cite{NSVZ_glue,sum_rules_for_scalar_glue,two_res_glue,narison98,shuryak_forkel,har00}.
Such a discrepancy would be indicative of 
two widely-separated  states, a result which has already been seen to occur in explicit two-resonance  
analyses of Laplace sum-rules even in  the absence of mixing with quark scalar resonances
\cite{two_res_glue,narison98}.
However, there exists substantial evidence  that these
discrepancies are resolved by the inclusion of instanton~\cite{basic_instanton}
effects in the Laplace sum-rules for scalar glueballs~\cite{shuryak_forkel,har00}.

Recently, techniques for using Gaussian sum-rules~\cite{gauss} to predict hadronic properties have been
developed~\cite{orl00}.  In particular, these methods concentrate on normalized Gaussian sum-rules which
are independent of the finite-energy sum-rule constraint which is central to
the original heat-evolution studies
\cite{gauss} of the Gaussian sum-rules.
Advantages of this approach compared with Laplace
sum-rules include enhanced sensitivity to the hadronic spectral function over a wide range of energy. 
In this paper, these techniques are employed and generalized in an effort to obtain resonance parameter predictions 
for scalar gluonium. Furthermore, the formulation of these sum-rules is extended to include Gaussian
kernels weighted by integer powers,
\begin{equation}
  \frac{1}{\sqrt{4\pi\tau}}\int\limits_{t_0}^\infty t^k
  \exp{\left[-\frac{(t-\hat s)^2}{4\tau}\right]} \frac{1}{\pi} \rho(t) \dif{t}
  \quad,\quad k\geq -1
\end{equation}
where $\rho(t)$ is a hadronic spectral function with physical threshold $t_0$.  
Similar to the Laplace sum-rules, the low-energy theorem (LET)~\cite{let}
(see~(\ref{let}) below) for scalar gluonic currents
enters {\em only} the $k=-1$ Gaussian sum-rule, and instanton contributions to the
correlation function serve to mitigate the discrepancy between the $k=-1$ and $k>-1$ sum-rules.
However, theoretical uncertainties associated with the instanton and LET parameters are
shown to be overwhelming in the $k=-1$ sum-rule, rendering it unsuitable for  phenomenological
analysis.
Thus the $k=0$ Gaussian sum-rule is the focus of our detailed predictions for
scalar glueballs, including an estimate of theoretical uncertainties.

We show that the Gaussian sum-rules of scalar gluonic currents contain signatures 
that the hadronic spectral function is distributed over a broad energy range,
(0.8--1.6) GeV including QCD uncertainties, and that this distribution most likely 
consists of two separate resonances. 
Since these sum-rules probe the gluonic content of hadronic states, 
they are sensitive to the glueball component of the observed scalar mesons
which in general could be glueball-quark meson mixtures.  
Thus our results are relevant to the interpretation of the scalar isoscalar resonances in this region:
the $f_0(400\mbox{--}1200)$, $f_0(980)$, $f_0(1370)$, $f_0(1500)$, and $f_0(1710)$ \cite{pdb}.

In the next section, Gaussian sum-rules for scalar gluonic currents are developed. 
In Section~\ref{analysis}, we develop analysis techniques and employ them to analyze the 
Gaussian sum-rules using a variety of phenomenological models. 
The results of this phenomenological analysis, including theoretical uncertainties,  
are consolidated in Section~\ref{discussion},
and a summary of our results is contained in Section~\ref{conclusion}.

\section{Scalar Glueball Gaussian Sum-Rules} \label{sumrules}
The most important quantity in any sum-rules approach to determining hadron properties is 
the correlation function for the particular channel under inspection:
\begin{equation}\label{corr}
  \Pi(Q^2)  = \ii\int\mathrm{d}^4\!x\, \e^{\ii q\cdot x} \langle\Omega | T\{J(x),J(0)\} |\Omega\rangle
  \quad,\quad Q^2=-q^2
\end{equation}
where $|\Omega\rangle$ is the QCD vacuum state, 
$T$ is the time-ordering operator, and $J(x)$ is that current which corresponds to the quantum numbers
of interest. In this paper, we wish to focus on scalar glueballs and so we choose the following
 current:
\begin{equation}\label{current}
  J = -\frac{\ppi^2}{\alpha\beta_0} \beta(\alpha)  G^a_{\ \mu\nu}  G^{a\mu\nu}
\end{equation}
which is renormalization-group invariant in the chiral limit of $n_{\text{f}}$ massless quarks.
The gluon field strength tensor $G^a_{\ \mu\nu}$ is defined by
\begin{equation}\label{gluefield}
  G^a_{\ \mu\nu} = \partial_{\mu} A^a_{\nu} - \partial_{\nu} A^a_{\mu}
  + g f^{abc} A^b_{\mu} A^c_{\nu}
\end{equation}
and $\beta(\alpha)$ is the QCD beta function describing the momentum 
scale dependence of the strong coupling parameter $\alpha$
\begin{gather}
  \beta\left(\alpha\right) =\nu^2\frac{d}{d\nu^2}\left(\frac{\alpha(\nu)}{\ppi}\right)=
  -\beta_0\left(\frac{\alpha}{\ppi}\right)^2-\beta_1\left(\frac{\alpha}{\ppi}\right)^3
  +\ldots
  \label{beta_expansion}\\
  \beta_0=\frac{11}{4}-\frac{1}{6} n_f\quad ,\quad
  \beta_1=\frac{51}{8}-\frac{19}{24}n_f   \quad,\quad\ldots\quad.
  \label{beta_coefficients}
\end{gather}

From the asymptotic form and assumed analytic properties of~(\ref{corr})
follows a dispersion relation with three subtraction constants
\begin{equation}\label{dispersion}
  \Pi(Q^2) - \Pi(0) - Q^2\Pi'(0) -\frac{1}{2}Q^4 \Pi''(0) =
     -\frac{Q^6}{\ppi} \int_{t_0}^{\infty} \frac{\rho(t)}{t^3 (t+Q^2)} \dif{t} \quad ,\quad Q^2 >0
\end{equation}
where $\rho(t)$ is the hadronic spectral function\footnote{In the literature, $\rho(t)$
is often denoted by $\Im\Pi(t)$.}  with physical threshold $t_0$.
The spectral function $\rho(t)$ is related to a physical process and is thus determined phenomenologically.
In contrast, $\Pi(Q^2)$  is calculated theoretically from QCD,
and the constant $\Pi(0)$ follows from the low-energy theorem~\cite{let}
\begin{equation}\label{let}
  \Pi(0)\equiv\lim_{Q^2\rightarrow 0} \Pi(Q^2) = \frac{8\pi}{\beta_0} \langle J\rangle\quad.
\end{equation}
For these reasons, we shall refer to the left-hand side of~(\ref{dispersion}) as the
theoretical side and the right-hand side as the phenomenological side.
In this regard, eqn.~(\ref{dispersion}) serves to relate theory to phenomenology, and, in principle,
could be used to predict the properties of hadrons from QCD.

However, as it stands, eqn.~(\ref{dispersion}) is not actually that well-suited to this task.
For instance, although the constant $\Pi(0)$ is determined by the low-energy theorem~(\ref{let}),
the constants $\Pi'(0)$ and $\Pi''(0)$ are not.
Further, the theoretical calculation of $\Pi(Q^2)$ contains a field theoretical
divergence proportional to $Q^4$.
In addition, from a phenomenological perspective, the integral on the right-hand side of~(\ref{dispersion})
is far too sensitive to the high energy behaviour of $\rho(t)$ to effectively probe low-lying
resonances.

To circumvent these shortcomings, we consider the one-parameter family of
Gaussian sum-rules\footnote{This definition is a natural generalization of that given in~\cite{gauss}.
To recover the original Gaussian sum-rule, we simply let $k=0$ in~(\ref{srdef}).}
\begin{equation}\label{srdef}
   G_k(\hat{s},\tau)\equiv \sqrt{\frac{\tau}{\ppi}}\borel
   \left\{ \frac{(\hat{s}+\ii\Delta)^k \Pi(-\hat{s}-\ii\Delta)
         - (\hat{s}-\ii\Delta)^k \Pi(-\hat{s}+\ii\Delta) }{\ii\Delta}
   \right\} \quad,\quad k\geq -1
\end{equation}
with the Borel transform $\borel$ defined by
\begin{equation}\label{borel}
  \borel\equiv \lim_{\stackrel{N,\Delta^2\rightarrow\infty}{\Delta^2/N\equiv 4\tau}}
  \frac{(-\Delta^2)^N}{\Gamma(N)}\left( \frac{\text{d}}{\text{d}\Delta^2}\right)^N \quad.
\end{equation}
Applying definition~(\ref{srdef})  to both sides of~(\ref{dispersion})
alleviates the difficulties surrounding~(\ref{dispersion}):
the infinite number of derivatives in~(\ref{borel}) annihilate the unwanted low-energy constants
and the field theoretical divergence contained in $\Pi(Q^2)$.
Furthermore, as we shall see, a key feature of the resulting sum-rules is the introduction of
a Gaussian weight factor to the integrand on the phenomenological side of~(\ref{dispersion}).
This serves to suppress contributions from  $\rho(t)$
away from the Gaussian peak---a desirable situation considering that we wish to extract
information concerning low-lying resonances.

Let us first consider~(\ref{srdef}) as applied to the theoretical side of~(\ref{dispersion}).
As noted previously,  the low-energy constants $\Pi'(0)$ and $\Pi''(0)$ are annihilated by
the Borel transform; however, the constant $\Pi(0)$ does produce a contribution unique to the
case $k=-1$. Using the following identity:
\begin{equation}\label{identity}
  \borel\left[ \frac{(\Delta^2)^n}{\Delta^2 +a}\right] =
  \frac{1}{4\tau} (-a)^n \exp\left( \frac{-a}{4\tau}\right)\ \mathrm{for}\ n\geq 0  \quad,
\end{equation}
it is trivial to show that the contribution to the theoretical side of the Gaussian sum-rules
devolving from the low-energy constant is given by
\begin{equation}\label{gauss_let}
  \frac{1}{\sqrt{4\pi\tau}} \exp\left(\frac{-\hat{s}^2}{4\tau} \right) \Pi(0) \quad.
\end{equation}

To proceed further, however,  we must settle on a specific form for the scalar glueball correlator $\Pi(Q^2)$.
We choose to partition the correlator into the following sum of qualitatively distinct terms:
\begin{equation}\label{gluecorr}
  \Pi^{\qcd}(Q^2) = \Pi^{\text{pert}}(Q^2)+\Pi^{\text{cond}}(Q^2)+\Pi^{\text{inst}}(Q^2)\quad,
\end{equation}
where the superscript \textit{QCD} signifies that~(\ref{gluecorr}) is a theoretical approximation
to the true correlator. The first two terms in~(\ref{gluecorr}) devolve from the operator
product expansion of the current~(\ref{current}).
The quantity $\Pi^{\text{pert}}(Q^2)$ is the contribution from ordinary perturbation
theory whereas $\Pi^{\text{cond}}(Q^2)$ is the result of nonzero vacuum expectation
values of local gluonic operators (condensates). For three colours and three flavours
of massless quarks ($n_{\text{f}}=3$), $\Pi^{\text{pert}}(Q^2)$ is given at three-loop
order by~\cite{che97}
\begin{gather}\label{pert}
  \Pi^{\text{pert}}(Q^2) = Q^4 \log\left(\frac{Q^2}{\nu^2}\right)
  \left[ a_0 + a_1 \log\left(\frac{Q^2}{\nu^2}\right) +a_2\log^2\left(\frac{Q^2}{\nu^2}\right)
  \right] \\
  a_0 = -2\left(\frac{\alpha}{\ppi}\right)^2\left[1+\frac{659}{36}\frac{\alpha}{\ppi}+
  247.480\left( \frac{\alpha}{\ppi}\right)^2\right] \, ,\quad
  a_1 = 2\left(\frac{\alpha}{\ppi}\right)^3\left[ \frac{9}{4}
            +65.781\frac{\alpha}{\ppi}\right] \, ,\quad
  a_2 = -10.1250\left(\frac{\alpha}{\ppi}\right)^4
  \label{pert_coefficients}
\end{gather}
where we have omitted the aforementioned field theoretical divergence
as this term is annihilated by the Borel operator~(\ref{borel}).
Incorporating into $\Pi^{\text{cond}}(Q^2)$ next-to-leading order~\cite{bag90}
contributions\footnote{The calculation of next-to-leading contributions in \cite{bag90} have been extended non-trivially to 
$n_f=3$ from $n_f=0$, and the operator basis has been changed from $\left\langle \alpha G^2\right\rangle$ to $\langle J\rangle$.} from the dimension four
gluon condensate $\langle J\rangle$ and leading order~\cite{NSVZ_glue}
contributions from gluonic condensates of dimension six and eight
\begin{gather}
   \label{dimsix}
   \left\langle {\cal O}_6\right\rangle  =
    \left\langle g f_{abc}G^a_{\mu\nu}G^b_{\nu\rho}G^c_{\rho\mu}\right\rangle \\
   \left\langle {\cal O}_8\right\rangle = 14\left\langle\left(\alpha f_{abc}G^a_{\mu\rho}
   G^b_{\nu\rho}\right)^2\right\rangle
   -\left\langle\left(\alpha f_{abc}G^a_{\mu\nu}G^b_{\rho\lambda}\right)^2\right\rangle
   \label{dimeight}
\end{gather}
yields
\begin{gather}\label{cond}
  \Pi^{\text{cond}}(Q^2) = \left[ b_0+b_1 \log\left( \frac{Q^2}{\nu^2} \right) \right]
  \langle J\rangle
  +\frac{c_0}{Q^2}\left\langle {\cal O}_6\right\rangle
  + \frac{d_0}{Q^4}\left\langle {\cal O}_8\right\rangle \\
  b_0 = 4\ppi\frac{\alpha}{\ppi}\left[ 1+ \frac{175}{36}\frac{\alpha}{\ppi}\right]
  \, ,\quad
  b_1 = -9\ppi\left(\frac{\alpha}{\ppi}\right)^2
  \, ,\quad
  c_0 = 8\ppi^2\left(\frac{\alpha}{\ppi}\right)^2
  \, ,\quad
  d_0 = 8\ppi^2\frac{\alpha}{\ppi} \quad.  \label{cond_coefficients}
\end{gather}
The final term on the right-hand side of~(\ref{gluecorr}) is a contribution
arising from direct instanton effects. Note that in decoupling this term from perturbation
theory and the condensate contributions, we have tacitly assumed that
interference between classical and quantum fields is small. Further, we also
assume that the dominant contribution to $\Pi^{\text{inst}}(Q^2)$ comes from BPST
single instanton and anti-instanton solutions~\cite{basic_instanton}
and that multi-instanton effects are negligible~\cite{schaefer_shuryak}.
With these provisions, we have~\cite{NSVZ_glue,shuryak_forkel,har00,inst_K2}
\begin{equation}\label{pi_inst}
\Pi^{\text{inst}}\left(Q^2\right)=
  32\ppi^2 Q^4\int \rho^4 \left[K_2\left(\rho\sqrt{Q^2}\right)\right]^2  \dif{n}(\rho)\quad ,
\end{equation}
where $K_2(x)$ is the modified Bessel function of the second kind of order two
(\emph{c.f.}~\cite{abr}), $\rho$ is the instanton radius, and $n(\rho)$
is the instanton density function.

\begin{figure}[htb]
\centering
\includegraphics[scale=0.5,angle=270]{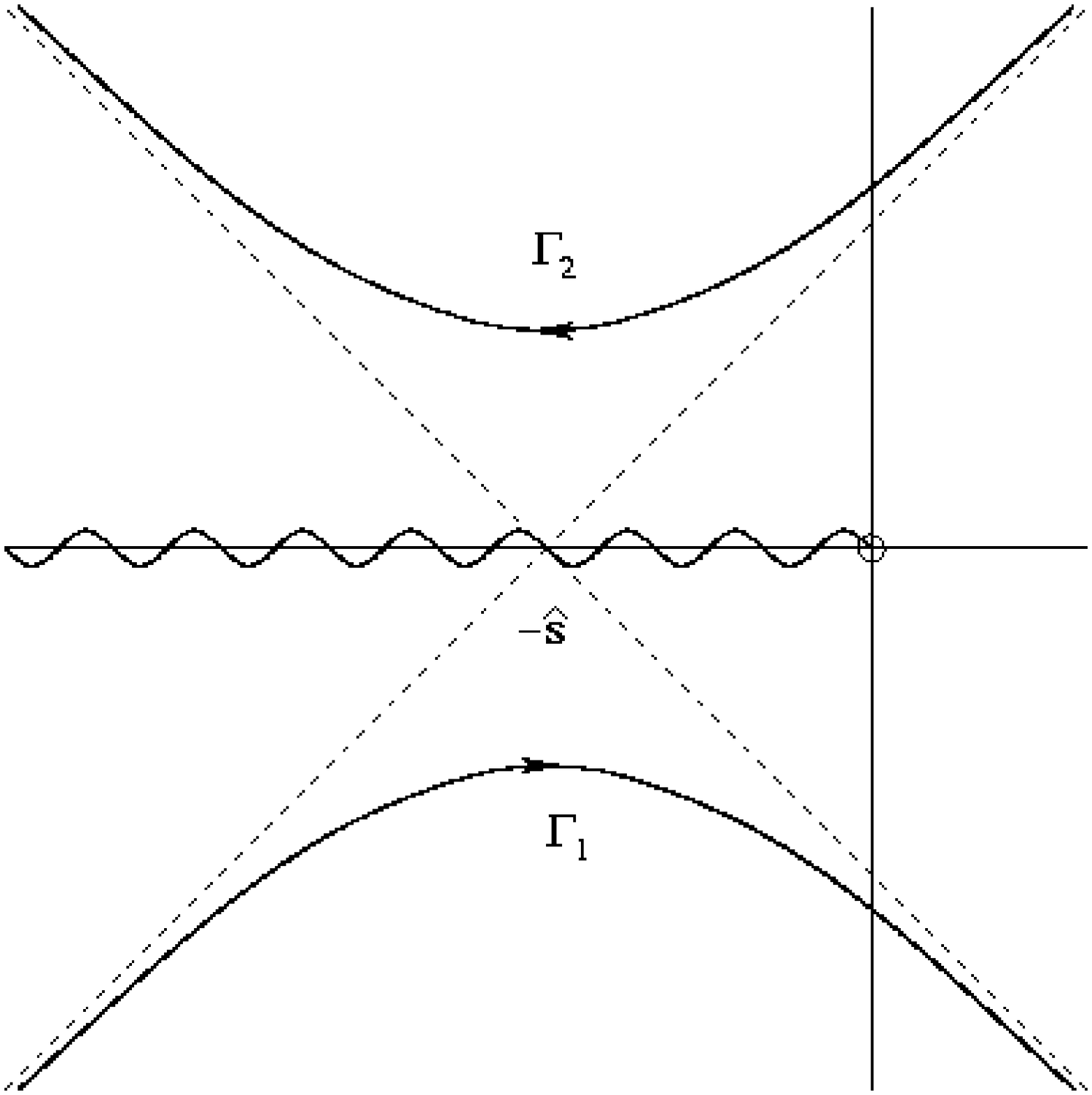}
\caption{Contour of integration $\Gamma_1+\Gamma_2$ defining the
Gaussian  sum-rule  in (\protect\ref{finish}).
The wavy line
on the negative real axis denotes the branch cut of $\Pi(w)$.
\label{cont_fig}}
\end{figure}

Before substituting~(\ref{gluecorr}) into~(\ref{srdef}), it is convenient to
first simplify~(\ref{srdef}) by employing a particularly useful identity relating the Borel
transform~(\ref{borel}) to the inverse Laplace transform~\cite{gauss}
\begin{equation}\label{bor_to_lap}
  \borel [f(\Delta^2)] = \frac{1}{4\tau}\il [f(\Delta^2)]
\end{equation}
where, in our notation,
\begin{equation}
  \il [f(\Delta^2)] = \frac{1}{2\ppi i} \int_{a-\ii\infty}^{a+\ii\infty}
    f(\Delta^2) \exp\left( \frac{\Delta^2}{4\tau} \right) \dif{\Delta^2}
\end{equation}
with $a$ chosen such that all singularities of $f$ lie to the left of $a$
in the complex $\Delta^2$-plane. (Note, there is no loss of generality in assuming
that $a>0$.) Rewriting~(\ref{srdef}) using~(\ref{bor_to_lap}) gives
\begin{multline} \label{mid}
   G_k(\hat{s},\tau) =
   \frac{1}{4\sqrt{\ppi\tau}}\frac{1}{2\ppi i}\left\{
   \int_{a-i\infty}^{a+i\infty}\frac{(\hat{s}+\ii\Delta)^k \Pi(-\hat{s}-\ii\Delta)}{\ii\Delta}
   \exp\left( \frac{\Delta^2}{4\tau} \right) \dif{\Delta^2}  \right.\\
   \left. -\int_{a-i\infty}^{a+i\infty}\frac{(\hat{s}-\ii\Delta)^k \Pi(-\hat{s}+\ii\Delta)}{\ii\Delta}
   \exp\left( \frac{\Delta^2}{4\tau} \right) \dif{\Delta^2}\right\} \quad.
\end{multline}
If in the first integral above, we make the substitution
$w=-\hat{s}-\ii\Delta$ and in the second, we make the substitution
$w=-\hat{s}+\ii\Delta$, then~(\ref{mid}) reduces to
\begin{align}
  G_k(\hat{s},\tau) & = \frac{1}{4\sqrt{\ppi\tau}}\frac{1}{2\ppi i}\left\{
   2\int_{\Gamma_1} (-w)^k \exp\left[ \frac{-(\hat{s}+w)^2}{4\tau}\right] \Pi(w)\dif{w}
  +2\int_{\Gamma_2} (-w)^k \exp\left[ \frac{-(\hat{s}+w)^2}{4\tau}\right] \Pi(w)\dif{w}\right\} \\
  \hspace{2em} & = \frac{1}{\sqrt{4\ppi\tau}}\frac{1}{2\ppi i}
  \int_{\Gamma_1 +\Gamma_2} (-w)^k  \exp
  \left[ \frac{-(\hat{s}+w)^2}{4\tau}\right] \Pi(w) \dif{w}
  \label{finish}
\end{align}
where $\Gamma_1$ and $\Gamma_2$ are two parabolas (depicted in Figure~\ref{cont_fig})
in the complex $w$-plane defined by
\begin{gather}
  \Gamma_1 =-\hat{s}-\ii(a^2+x^2)^{1/4} \exp\left[ \frac{\ii}{2}\atan
  \left( \frac{x}{a}\right) \right] \label{gamma1} \\
  \Gamma_1 =-\hat{s}+\ii(a^2+x^2)^{1/4} \exp\left[ \frac{\ii}{2}\atan
  \left( \frac{x}{a}\right) \right] \label{gamma2}
\end{gather}
for all $x\in\mathbb{R}$.

\begin{figure}[htb]
\centering
\includegraphics[scale=0.5,angle=270]{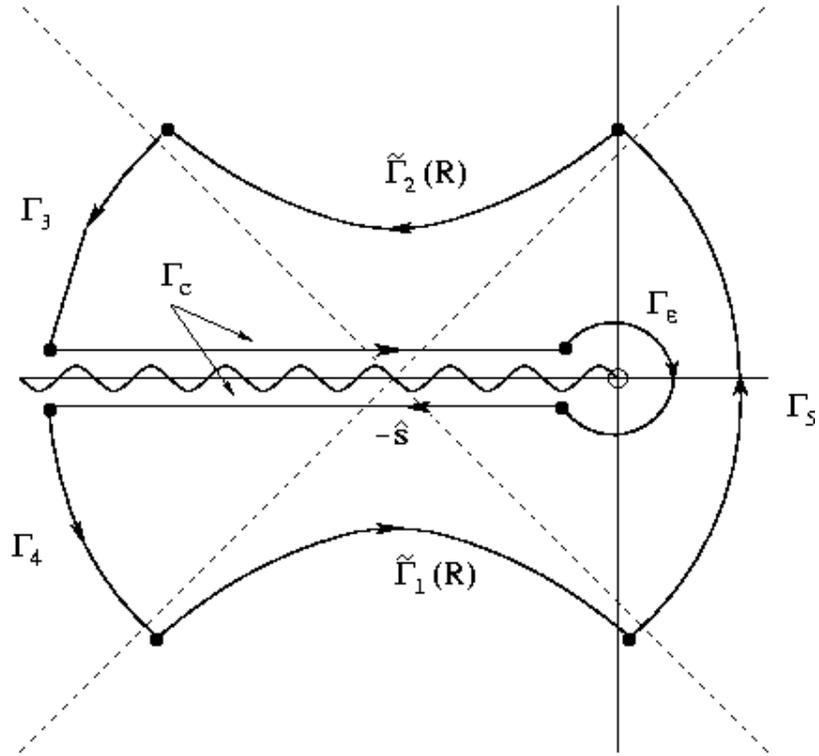}
\caption{Closed contour $C(R)$ used to
calculate the Gaussian sum-rule defined by  (\protect\ref{finish}).
The  inner circular segment $\Gamma_\epsilon$ has a radius of $\epsilon$, and the  circular
segments
$\Gamma_3$, $\Gamma_4$ and $\Gamma_5$
have a radius $R$.
The wavy line
on the negative real axis denotes the branch cut of $\Pi(w)$,
and the linear segments of the contour above and below the branch cut are denoted by $\Gamma_c$.
The contour $\tilde{\Gamma}_1(R)$ is that portion of $\Gamma_1$ (see Figure \protect\ref{cont_fig})
which lies in the interior of a circle of radius $R$ centered at $-\hat s$, and similarly
for  $\tilde{\Gamma}_2(R)$.
\label{deformed}}
\end{figure}

Now, we must substitute~(\ref{gluecorr}) into~(\ref{finish}) and calculate the resulting complicated integral.
Towards this end, it is advantageous to
consider the closed contour $C(R)$ depicted in Figure~\ref{deformed}. Our
expression for the correlator~(\ref{gluecorr}) is analytic in the complex $w\equiv Q^2$-plane except
for a branch cut along the negative real semi-axis originating from a branch point
located at the origin. Consequently,
\begin{align}
 0 & = \frac{1}{2\ppi i}\frac{1}{\sqrt{4\ppi\tau}} \oint_{C(R)} (-w)^k  \exp \left[
         \frac{-(\hat{s}+w)^2}{4\tau} \right] \Pi^{\qcd}(w)\dif{w} \\
    & = \frac{1}{2\ppi i}\frac{1}{\sqrt{4\ppi\tau}}
     \left\{ \int_{\tilde{\Gamma}_1(R)+\tilde{\Gamma}_2(R)}
           +\quad\int_{\Gamma_{\text{c}}+\Gamma_{\epsilon}}
           +\quad\int_{\Gamma_3+\Gamma_4+\Gamma_5} \right\}
        (-w)^k \exp \left[ \frac{-(\hat{s}+w)^2}{4\tau} \right] \Pi^{\qcd}(w)\dif{w}
         \label{big}
\end{align}
where $\tilde{\Gamma}_1(R)$ and $\tilde{\Gamma}_2(R)$ are respectively those
portions of the contours $\Gamma_1$ and $\Gamma_2$ (see~(\ref{gamma1}) and~(\ref{gamma2}))
lying in the interior of a circle of radius $R$ centered at $-\hat{s}$.
For large $R$, the integral over $\Gamma_3+\Gamma_4+\Gamma_5$ approaches
zero and the contours $\tilde{\Gamma}_1(R)$ and $\tilde{\Gamma}_2(R)$ approach
$\Gamma_1$ and $\Gamma_2$.
Therefore, by rearranging~(\ref{big}), recalling~(\ref{finish}), and taking appropriate
limits,  we get
\begin{align}
  G_k^{\qcd} (\hat{s},\tau)  &= -\frac{1}{2\pi\ii}\frac{1}{\sqrt{4\ppi\tau}}
  \lim_{\stackrel{R\rightarrow\infty}{\epsilon\rightarrow 0} }
  \int_{\Gamma_{\text{c}}+\Gamma_{\epsilon}}
  (-w)^k  \exp \left[ \frac{-(\hat{s}+w)^2}{4\tau} \right] \Pi^{\qcd}(w)\dif{w} \\
  &= \frac{1}{2\ppi\ii}\frac{(-1)^{k+1}}{\sqrt{4\ppi\tau}}  \lim_{\epsilon\rightarrow 0}
        \int_{\Gamma_{\epsilon}} w^k
        \exp\left[ \frac{-(\hat{s}+w)^2}{4\tau} \right] \Pi^{\qcd}(w) \dif{w} \nonumber\\
   &\hspace{10em}+ \frac{1}{\sqrt{4\ppi\tau}} \lim_{\epsilon\rightarrow 0}
        \int_{\epsilon}^{\infty}  t^k
        \exp \left[ \frac{-(\hat{s}-t)^2}{4\tau} \right] \frac{1}{\ppi}\Im\Pi^{\qcd}(t) \dif{t} \quad,
        \label{theorysr}
\end{align}
where
\begin{equation}
   \Im\Pi^{\qcd}(t) \equiv  \lim_{\delta\rightarrow 0^+}
  \left[ \frac{\Pi^{\qcd}(-t-\ii\delta)-\Pi^{\qcd}(-t+\ii\delta)}{2\ii}\right]\quad.
\end{equation}
Eqn.~(\ref{theorysr}) is our final expression for the contribution to the $k$-th Gaussian sum-rule
of scalar gluonium stemming from the correlator~(\ref{gluecorr}).
Later in this section, however, we do evaluate~(\ref{theorysr}) for the specific cases
$k=-1,0$ (see~(\ref{GM1}) and~(\ref{G0})).

We must now consider the phenomenological contribution to the Gaussian sum-rules.
Substituting the right-hand side of~(\ref{dispersion}) into~(\ref{srdef}) and again making use
of the identity~(\ref{identity}), it is simple to show that
\begin{equation}\label{phenom_sr}
  G_k^{\text{phen}} (\hat{s},\tau) =\frac{1}{\sqrt{4\ppi\tau}}\int_{t_0}^{\infty}
   t^k \exp\left[ \frac{-(\hat{s}-t)^2}{4\tau}\right] \frac{1}{\ppi}\rho(t) \dif{t}  \quad.
\end{equation}
In sum-rules analyses,
it is customary to approximate the spectral function $\rho(t)$ using a  ``resonance(s) + continuum''
ansatz. In this model, hadronic physics is (locally) dual to QCD above the continuum threshold $s_0$,
and so we write
\begin{equation}\label{phenom}
  \rho(t)=\rho^{\text{had}}(t)+\theta\left(t-s_0\right){\rm Im}\Pi^{\qcd}(t)
\end{equation}
where  $\theta (t)$ is the Heaviside step function.
(We shall have much more to say concerning $\rho^{\text{had}}(t)$  in Section~\ref{analysis}.)

Substituting~(\ref{phenom}) into~(\ref{phenom_sr}) and comparing the result to
the theoretical expression~(\ref{theorysr}) shows us that the continuum contribution
\begin{equation}\label{continuum}
   G_k^{\text{cont}} (\hat{s},\tau,s_0) = \frac{1}{\sqrt{4\pi\tau}}   \int_{s_0}^{\infty} t^k
   \exp \left[ \frac{-(\hat{s}-t)^2}{4\tau} \right]  \frac{1}{\pi} \Im \Pi^{\qcd}(t) \dif{t}
\end{equation}
is common to both; therefore, we define
\begin{gather}
  G_k^{\qcd}(\hat{s},\tau,s_0) \equiv G_k^{\qcd}(\hat{s},\tau) -  G_k^{\text{cont}} (\hat{s},\tau,s_0) \label{blah} \\
  G_k^{\text{had}}(\hat{s},\tau) \equiv G_k^{\text{phen}}(\hat{s},\tau) -  G_k^{\text{cont}} (\hat{s},\tau,s_0)
\end{gather}
and write  (recall the low-energy contribution~(\ref{gauss_let}) unique to the $k=-1$ sum-rule)
\begin{gather}
  G_{-1}^{\qcd}(\hat{s},\tau,s_0)  + \frac{1}{\sqrt{4\pi\tau}} \exp\left( \frac{-\hat{s}^2}{4\tau}\right)
   \Pi(0) =  G_{-1}^{\text{had}}(\hat{s},\tau)\label{srm1}  \\
  G_k^{\qcd}(\hat{s},\tau,s_0) = G_k^{\text{had}}(\hat{s},\tau)\quad,\quad k\geq 0
  \label{sr0}
\end{gather}
with
\begin{equation}
   G_k^{\text{had}}(\hat{s},\tau) = \frac{1}{\sqrt{4\pi\tau}} \int_{t_0}^{\infty} t^k
    \exp\left[ \frac{-(\hat{s}-t)^2}{4\tau}\right] \frac{1}{\pi} \rho^{\text{had}}(t) \dif{t}\quad,\quad
    k\geq-1\quad.
\label{G_had_def}
\end{equation}
We note that~(\ref{srm1}--\ref{G_had_def}) have meaning for $\hat s <0$ since
this simply represents a Gaussian kernel whose peak lies outside the $t>0$
physical region, and hence only the Gaussian tail extends into the physical
region.  As will later be shown, the QCD results scale with $\tau$ through the 
renormalization-group equation, and hence QCD also presents no obstacles to
considering $\hat s<0$.  Indeed, the seminal work \cite{gauss} on Gaussian 
sum-rules considered symmetric and antisymmetric combinations 
$U^{\pm}\left(\hat s,\tau\right)=G\left(\hat s, \tau\right)\pm G\left(-\hat s,
  \tau\right)$
which implicitly employs Gaussian sum-rules with negative $\hat s$.

In this paper, we focus exclusively on the $k=0,-1$ Gaussian sum-rules. Substituting~(\ref{gluecorr})
into~(\ref{blah}) and recalling~(\ref{theorysr}) gives\footnote{The given result is 
valid to leading order in the condensates.}
(for details on simplifying the relevant integrals in~(\ref{theorysr}), see~\cite{har00,orl00})
\begin{equation}\label{GM1}
\begin{split}
     G_{-1}^{\qcd}(\hat{s},\tau,s_0) = &-\frac{1}{\sqrt{4\pi\tau}} \int_0^{s_0} t
     \exp\left[ \frac{-(\hat{s}-t)^2}{4\tau}\right]  \left[ (a_0-\pi^2 a_2)
     +2a_1\log\left( \frac{t}{\nu^2} \right)  + 3a_2 \log^2\left( \frac{t}{\nu^2} \right) \right]\dif{t}
   \\
      &+ \frac{1}{\sqrt{4\pi\tau}} \exp\left( \frac{-\hat{s}^2}{4\tau}\right)
       \left[ -b_0 \langle J\rangle + \frac{c_0 \hat{s}}{2\tau}\left\langle {\cal O}_6\right\rangle
      -\frac{d_0}{4\tau}\left(\frac{\hat{s}^2}{2\tau} -1 \right)
      \left\langle {\cal O}_8\right\rangle \right]
   \\
      &-\frac{16\pi^3}{\sqrt{4\pi\tau}} \int\dif{n}(\rho)\rho^4
      \int_0^{s_0} t  \exp\left[ \frac{-(\hat{s}-t)^2}{4\tau}\right]  J_2\left(\rho\sqrt{t} \right)
      Y_2\left(\rho\sqrt{t} \right)  \dif{t}
   \\
      &-\frac{128\pi^2}{\sqrt{4\pi\tau}} \exp\left( \frac{-\hat{s}^2}{4\tau}\right)
      \int\dif{n}(\rho)
\end{split}
\end{equation}
\begin{equation}\label{G0}
\begin{split}
      G_0^{\qcd}(\hat{s},\tau,s_0) = &- \frac{1}{\sqrt{4\pi\tau}} \int_0^{s_0} t^2
       \exp\left[ \frac{-(\hat{s}-t)^2}{4\tau}\right]  \left[ (a_0-\pi^2 a_2)
       +2a_1\log\left( \frac{t}{\nu^2} \right)  + 3a_2 \log^2\left( \frac{t}{\nu^2} \right) \right]
       \dif{t}
   \\
       &- \frac{1}{\sqrt{4\pi\tau}} b_1\langle J\rangle \int_0^{s_0}
        \exp\left[ \frac{-(\hat{s}-t)^2}{4\tau}\right] \dif{t}
        + \frac{1}{\sqrt{4\pi\tau}} \exp\left( \frac{-\hat{s}^2}{4\tau}\right)
         \left[ c_0 \left\langle {\cal O}_6\right\rangle  - \frac{d_0 \hat{s}}{2\tau}
         \left\langle {\cal O}_8\right\rangle \right]
   \\
       &-\frac{16\pi^3}{\sqrt{4\pi\tau}} \int\dif{n}(\rho)\rho^4
       \int_0^{s_0} t^2  \exp\left[ \frac{-(\hat{s}-t)^2}{4\tau}\right]  J_2\left(\rho\sqrt{t} \right)
       Y_2\left(\rho\sqrt{t} \right)  \dif{t}
\end{split}
\end{equation}
where $J_2(x)$ and $Y_2(x)$ are Bessel functions of order two of the first and second kind
respectively (\textit{c.f.}~\cite{abr}).
Renormalization-group improvements~\cite{gauss,nar81} of~(\ref{GM1}) and~(\ref{G0}) amount to replacing the
strong coupling constant $\alpha$ (contained in the coefficients~(\ref{pert_coefficients})
and~(\ref{cond_coefficients})) by the running coupling $\alpha(\nu^2)$ at the renormalization scale
$\nu^2=\sqrt{\tau}$.  At three loop order with $n_{\text{f}}=3$  in the $\overline{\text{MS}}$
renormalization scheme, we have~\cite{pdb}
\begin{gather}
  \frac{\alpha (\nu^2)}{\pi} = \frac{1}{\beta_0 L}-\frac{\bar\beta_1\log L}{\left(\beta_0L\right)^2}+
  \frac{1}{\left(\beta_0 L\right)^3}\left[
  \bar\beta_1^2\left(\log^2 L-\log L -1\right) +\bar\beta_2\right]
  \label{run_coupling}\\
  L=\log\left(\frac{\nu^2}{\Lambda^2}\right)\quad ,\quad \bar\beta_i=\frac{\beta_i}{\beta_0}
  \quad ,\quad
  \beta_0=\frac{9}{4}\quad ,\quad \beta_1=4\quad ,\quad \beta_2=\frac{3863}{384}
\end{gather}
with
 $\Lambda_{\overline{MS}}\approx 300\,{\rm MeV}$ for three active flavours,
consistent with current estimates of $\alpha(M_\tau)$~\cite{pdb}
\footnote{The role of higher-loop effects on
 the extraction of $\alpha\left(M_\tau\right)$ and the subsequent impact on
 $\Lambda_{\overline{MS}}$ has been investigated through Pad\'e approximation
 techniques \protect\cite{ste98}.} and
matching conditions through the charm threshold~\cite{che98}.

The normalization of  the Gaussian sum-rules
is related to the finite-energy sum-rules (FESRs)~\cite{fesr}
as can be seen by integrating~(\ref{srm1}) and~(\ref{sr0})  with respect to $\hat{s}$ to obtain
\begin{gather}
   \int\limits_{-\infty}^\infty G_{-1}^{\qcd}(\hat{s}, \tau,s_0) \dif{\hat{s}}   + \Pi(0)
  =\frac{1}{\pi}\int\limits_{t_0}^{\infty} \frac{1}{t}\rho^{\text{had}}(t) \dif{t}
  \label{tom_norm_1} \\
  \int\limits_{-\infty}^\infty G_k^{\qcd}(\hat{s}, \tau,s_0) \dif{\hat{s}}
  =\frac{1}{\pi}\int\limits_{t_0}^{\infty} t^k\rho^{\text{had}}(t) \dif{t} \quad,\quad k\geq 0.
  \label{tom_norm_2}
\end{gather}
We recognize the quantities on the right-hand sides of (\ref{tom_norm_1})  and~(\ref{tom_norm_2})
from the definition of the (FESRs)
\begin{equation}
F_k\left(s_0\right)=\frac{1}{\pi}
\int\limits_{t_0}^{\infty}\mathrm{d}t\, t^k\rho^{\text{had}}(t)\quad ,
\label{tom_fesr}
\end{equation}
where $F_k$ represents a QCD prediction.  Thus we see that the overall normalization of
$G_k^{\qcd}$ (or $G_{-1}^{\qcd}+\Pi(0)$) is constrained by the finite-energy sum-rules.

This result is not surprising in light of the seminal work on Gaussian sum-rules
which established the significance of the FESR constraint by considering the evolution of the Gaussian
sum-rules through the diffusion equation~\cite{gauss}.  It was found that this
``heat-evolution'' of the resonance plus continuum  model would  only reproduce the QCD prediction in the asymptotic
regime if the continuum $s_0$ was constrained by the lowest FESR.  Hence, the normalization of the Gaussian
sum-rules, which is constrained by the FESR, should be removed by defining normalized (unit-area)
Gaussian sum-rules
\begin{gather}
  N^{\qcd}_{-1} (\hat{s}, \tau, s_0) \equiv
  \frac{ G^{\qcd}_{-1} (\hat s, \tau, s_0) + \frac{1}{\sqrt{4\pi\tau}}\exp
    \left(\frac{-\hat{s}^2}{4\tau} \right)\Pi(0)}{M^{\qcd}_{-1,0} (\tau, s_0)+\Pi(0)}
  \label{tom_norm_srm1} \\
  N^{\qcd}_k (\hat{s}, \tau, s_0) \equiv
  \frac{G^{\qcd}_{k} (\hat s, \tau, s_0)}{M^{\qcd}_{k,0} (\tau, s_0)} \quad, \quad k\geq 0
  \label{tom_norm_srk}
\end{gather}
where the $n$-th moment of $G_k$ is given by
\begin{equation}\label{moments}
  M_{k,n}(\tau, s_0)
  = \int\limits_{-\infty}^\infty \hat{s}^n G_k (\hat s,\tau, s_0) \dif{\hat{s}}
  \quad,\quad n=0,1,2,\ldots\quad.
\end{equation}
Note that, for the sake of notational convenience in subsequent sections, we have
absorbed the low-energy theorem contribution into the definition of $N^{\qcd}_{-1}$
(see~(\ref{tom_norm_srm1}) above).
This allows us to write our final version of the normalized Gaussian sum-rules
of scalar gluonium as
\begin{equation}\label{ngsr}
   N_k^{\qcd}(\hat{s},\tau,s_0) = \frac{ \frac{1}{\sqrt{4\pi\tau}} \int_{t_0}^{\infty} t^k
   \exp\left[\frac{-(\hat{s}-t)^2}{4\tau} \right] \frac{1}{\pi}\rho^{\text{had}}(t)\dif{t}}{\int_{t_0}^{\infty} t^k
   \frac{1}{\pi}\rho^{\text{had}}(t)} \quad,\quad k\geq -1\quad.
\end{equation}

Before describing analysis methods for the  Gaussian sum-rules,
we  specify the numerical values for the QCD parameters appearing in~(\ref{gluecorr}) that will be employed in our analysis.
For the dimension four gluon condensate, we make the assumption that
$\langle J\rangle\approx\langle\alpha G^a_{\ \mu\nu}G^{a\mu\nu} \rangle$
and then employ the most recently updated value~\cite{nar97}
\begin{equation}\label{dimfour}
   \langle\alpha G^a_{\ \mu\nu}G^{a\mu\nu}\rangle
      = (0.07\pm 0.01)\ \unit{GeV}^4 \quad.
\end{equation}
The dimension six gluon condensate can be related to
$\langle\alpha G^a_{\ \mu\nu}G^{a\mu\nu}\rangle$ using
instanton techniques (see~\cite{NSVZ_glue,SR})
\begin{equation}
   \langle \mathcal{O}_6 \rangle = (0.27\ \unit{GeV}^2)
   \langle \alpha G^a_{\ \mu\nu}G^{a\mu\nu}\rangle\quad.
\end{equation}
Further, by invoking vacuum saturation in conjunction with the heavy quark
expansion, the authors of~\cite{bag85} have related the dimension eight
gluon condensate to $\langle\alpha G^a_{\ \mu\nu}G^{a\mu\nu}\rangle$
through
\begin{equation}
   \langle\mathcal{O}_8\rangle = \frac{9}{16}
   \left(\langle\alpha G^a_{\ \mu\nu}G^{a\mu\nu}\rangle\right)^2\quad.
\end{equation}
Regarding the instanton contributions, we shall employ Shuryak's dilute instanton
liquid model~\cite{DIL} in which
\begin{equation}
   n(\rho) = n_{\text{c}} \delta(\rho-\rho_{\text{c}})
\end{equation}
with
\begin{equation} \label{DILparams}
  n_{\text{c}} = 8.0\times 10^{-4}\ \unit{GeV}^4\ \mathrm{and}\ \rho_{\text{c}} =
  \frac{1}{0.6}\ \unit{GeV}^{-1}\quad.
\end{equation}

\section{Analysis of the Scalar Glueball Gaussian Sum-Rules}\label{analysis}
In most sum-rules analyses, it is necessary to make some assumptions concerning
the hadronic content of the spectral function~(\ref{phenom}).  
A common choice in the literature is that of a single narrow resonance model. 
However, the wealth of scalar states with masses below $2\ \unit{GeV}$~\cite{pdb}, 
some of which are quite broad, certainly raises a question as to the validity of 
such an assumption, and suggest that  a more general model allowing for a distribution of the resonance 
strength would be more suitable. 
In the following subsections, we show that this is indeed the case. 
Using appropriate generalizations of the analysis techniques developed in~\cite{orl00},
we demonstrate that the single narrow resonance model provides an
inadequate description of the hadronic content of the scalar glueball correlator, 
whereas certain  distributed strength models lead to outstanding
agreement between the theory and phenomenology parameterized by~(\ref{ngsr}).

\subsection{Single Narrow Resonance Model}\label{snrm}
In the single narrow resonance model, $\rho^{\text{had}}(t)$ takes the form
\begin{equation}\label{single}
  \rho^{\text{had}}(t) = \pi f^2 \delta(t-m^2)
\end{equation}
where $m$ and $f$ are respectively the resonance mass and coupling. With such an ansatz, the
normalized Gaussian sum-rule~(\ref{ngsr}) becomes
\begin{equation}\label{phenom_single}
  N_k^{\qcd}(\hat{s},\tau,s_0) = \frac{1}{\sqrt{4\ppi\tau}}
  \exp\left[ -\frac{(\hat{s}-m^2)^2}{4\tau}\right]
  \quad.
\end{equation}

The phenomenological side of~(\ref{phenom_single}) admits an absolute maximum (peak) 
located at $\hat{s}=m^2$, \textit{independent} of $\tau$; therefore,
the theoretical side of~(\ref{phenom_single}) should mimic this behaviour
provided that the single narrow resonance model~(\ref{single}) is actually an adequate
description of hadronic physics below the continuum threshold 
({\it i.e.}\ heavier states are weakly coupled enough to be absorbed into the continuum).
Defining $\hat{s}_{\text{peak}}(\tau, s_0)$ by the condition 
\begin{equation}\label{peakpos}
  \frac{\partial}{\partial \hat{s}} N^{\text{QCD}}_k
  (\hat{s}_{\text{peak}}(\tau, s_0),\tau,s_0)= 0 \quad,
\end{equation} 
and denoting by $\{\tau_n\}_{n=0}^N$ an equally spaced partition of the
$\tau$ interval of interest $[\tau_{\text{i}},\tau_{\text{f}}]$  (we
elaborate on this interval shortly), we define the following $\chi^2$-function
\begin{equation}\label{chi_single}
  \chi^2(s_0,m^2) = \sum_{n=0}^{N} \left[ \hat{s}_{\text{peak}}(\tau_n, s_0)
  - m^2 \right] ^2  
\end{equation} 
as a measure of the difference between the theoretical peak position and the
phenomenological peak position.
Minimization of~(\ref{chi_single}) with respect to $s_0$ and $m^2$
then provides us with values for these two parameters which correspond to
the best possible fit between the theory and phenomenology as represented
through~(\ref{phenom_single}).
Lastly, the optimizing condition
\begin{equation}
  \frac{\partial}{\partial m^2} \chi^2(s_0,m^2) =0
\end{equation}
allows us to write $m^2$ as a function of $s_0$ whereby reducing~(\ref{chi_single}) to 
a one-dimensional minimization problem:
\begin{equation}\label{final_single_chi}
\chi^2(s_0) = \sum_{n=0}^{N} \left[ \hat{s}_{\text{peak}}(\tau_n, s_0)
  - m^2(s_0) \right] ^2 
\end{equation}
with
\begin{equation}\label{single_mass}
  m^2(s_0) = \frac{1}{N+1} \sum_{n=0}^N \hat{s}_{\text{peak}}(\tau_n, s_0)\quad. 
\end{equation}
Thus, in a single narrow resonance analysis,
we first minimize~(\ref{final_single_chi}) with respect to $s_0$
to determine an optimum choice for the continuum threshold parameter and
then substitute this value into~(\ref{single_mass}) to obtain the best fit
resonance mass. 

Upon obtaining these optimized parameters,
there exists criteria that can be used to assess the validity 
of this phenomenological model.
For instance, if the single resonance analysis is a reasonable approach,
plots of the theoretical Gaussian sum-rules
(the left-hand side of~(\ref{phenom_single})) and plots of
the phenomenological Gaussian sum-rules (the right-hand side of~(\ref{phenom_single}))
should coincide (to a large degree). Significant deviation of one from the other
may be indicative of an inadequate phenomenological model.
On a more quantitative note, consider the following 
combination of moments~(\ref{moments})
(where we suppress the explicit dependence on $\tau$ and $s_0$):
\begin{equation}\label{sigma_combo}
  \sigma^2_k \equiv \frac{M_{k,2}}{M_{k,0}} -\left(\frac{M_{k,1}}{M_{k,0}}\right)^2 
\end{equation}
In the single narrow resonance model we should find
\begin{equation}
  \sigma_k^2-2\tau = 0 \quad, 
\label{b}
\end{equation}
and hence a substantial deviation of $\sigma_k^2-2\tau$ from zero signals a failure of the 
single narrow resonance model.

To proceed with the analysis, however, 
we must first choose our region of interest $[\tau_{\text{i}}, \tau_{\text{f}}]$
needed in the definition of the $\chi^2$ function~(\ref{chi_single}).
There are a number of factors to be considered in selecting this interval.
The lower bound $\tau_{\text{i}}$ must be large enough such that the condensate contributions
do not dominate perturbation theory and also such that the leading omitted perturbative
term in the expansion for the running coupling~(\ref{run_coupling}) is small.
Therefore, in accordance with these criteria, we choose a lower bound of
$\tau_{\text{i}}\geq 2\ \unit{GeV}^4.$
To choose an appropriate upper bound on $\tau$, we first note that the Gaussian kernel
has a resolution of $\sqrt{2\tau}$. It is important to the analysis that
the Gaussian sum-rules employed have a resolution
less than the non-perturbative (hadronic physics) energy scale involved:
roughly 2--3 $\unit{GeV}^2$.  This fact motivates an upper bound of $\tau_{\text{f}}\leq
4\ \unit{GeV}^4$.
Therefore, \emph{in this and in all subsequent analyses}, we restrict our attention to the range
$2\ \unit{GeV}^4\leq\tau\leq\ 4\ \unit{GeV}^4$.

An analysis of the $k=-1$ normalized Gaussian sum-rule through~(\ref{phenom_single})
leads to predictions that are completely unstable under QCD uncertainties.
Incorporating the error bounds of the dimension four gluonic condensate~(\ref{dimfour})
and allowing for a 15\% error in each of $n_{\text{c}}$ and $\rho_{\text{c}}$ leads to
a huge degree of variation in the resulting hadronic parameter estimates, and
it is not even possible to ascertain whether or not an extension of 
the single resonance analysis is warranted.
Mass predictions range anywhere from $1.0\ \unit{GeV}$ to $1.8\ \unit{GeV}$,
an interval far too broad to be of much use.
The only inference we can draw from $k=-1$ analysis is that the mass scale
obtained is in rough agreement with that which results from an analysis
of the next-to-leading order ($k=0$) Gaussian sum-rule (see below).
We note further that this consistency in mass scales between the $k=-1$ and higher order
sum-rules is also observed in~\cite{shuryak_forkel,har00} and occurs only when instanton
effects are included.
Due to its extreme sensitivity to small variations in various QCD parameters,
we are forced to conclude that the $k=-1$ Gaussian sum-rule is an unreliable probe
of the scalar glueball sector, and so, move on to a $k=0$ analysis. 

A single narrow resonance analysis of the $k=0$ Gaussian sum-rule,
however, leads to rather poor agreement between theory and phenomenology.
Minimization of~(\ref{final_single_chi})
leads to an optimum threshold at $s_0=2.3\ \unit{GeV}^2$ which, when substituted into~(\ref{single_mass}),
yields a resonance mass of $m=1.30\ \unit{GeV}$. In Figure~\ref{sigma}, we plot both
$\sigma_0^2(\tau,s_0)$ and $2\tau$ versus $\tau$ for $s_0=2.3\ \unit{GeV}^2$.
The graph of $\sigma_0^2(\tau,s_0)$ appears to have a slope of two, but, if extended
to $\tau=0$, would not pass through the origin---a situation which contradicts
the single resonance result~(\ref{b}).
Further, in Figure~\ref{sum_rules_single}, we plot
the left- (theoretical) and right- (phenomenological) hand sides of~(\ref{phenom_single})
versus $\hat{s}$ for $\tau\in\{2,3,4\}\ \unit{GeV}^4$  using the optimized values
$m=1.30\ \unit{GeV}$ and $s_0=2.3\ \unit{GeV}^2$.
The discrepancy between theory and phenomenology is apparent: the theoretical curves
consistently underestimate phenomenology near the peak and overestimate
phenomenology in the tails.  These observations indicate
that a single narrow resonance ansatz is an inadequate description of the scalar gluonium
spectral function, hence motivating our subsequent analyses of more general models. 

\subsection{Distributed Resonance Strength Models}\label{distributed}
The moment combination $\sigma_k^2$ measures the width of the QCD distribution, and  
since $\sigma_0^2-2\tau>0$,
we conclude that the resonance strength is distributed over an energy region 
broad enough to be sampled by the resolution $\sqrt{2\tau}$ of the Gaussian kernel.  
In accordance, we should therefore consider phenomenological models which allow for 
such a spreading. We focus on three distinct classes of distributed strength models:
\begin{enumerate}
  \item single non-zero width models,
  \item a double narrow resonance model, and
  \item single narrow resonance + single non-zero width resonance models,
\end{enumerate}
ordered according to increasing complexity, \textit{i.e.} the (normalized) single
non-zero width models each contain two free parameters whereas the single narrow
resonance + single non-zero width models requires four. 
In the subsections to follow, we outline the analysis procedure and the resulting 
hadronic parameter predications corresponding to each of the models under consideration. 

\subsubsection{Single Non-Zero Width Models}\label{sfwr}
The first non-zero width model we consider is a unit-area square pulse
representing a broad, structureless background which has
previously been used for the study of the $\sigma$ resonance~\cite{Elias}.
To describe such a square pulse centred at $m^2$ and having width $2m\Gamma$, we define
\begin{equation}
\frac{1}{\pi}\rho^{\text{had}}(t)=
\frac{1}{\pi}\rho^{sp}(t)\equiv
\frac{1}{2m\Gamma}\left[
\theta\left(t-m^2+m\Gamma\right)-\theta\left(t-m^2-m\Gamma\right)
\right]\quad ,
\label{square_pulse}
\end{equation}
which leads to the following normalized Gaussian sum-rule~(\ref{ngsr}) for $k=0$:
\begin{equation}
 N^{\qcd}_0(\hat{s},\tau,s_0)
=\frac{1}{4m\Gamma}\left[
{\rm erf}\left(\frac{\hat s-m^2+m\Gamma}{2\sqrt{\tau}}\right)
-{\rm erf}\left(\frac{\hat s-m^2-m\Gamma}{2\sqrt{\tau}}
\right)
\right]\quad .
\label{gauss_sp}
\end{equation}

As in the single narrow resonance analysis, the phenomenological side of~(\ref{gauss_sp})
admits a single $\tau$-independent peak, and so we can determine an optimum threshold parameter
$s_0$ by following the steps outlined through~(\ref{chi_single})--(\ref{single_mass})
\textit{i.e.} $s_0=2.3\ \text{GeV}^2$.
We wish to use this optimized value of $s_0$ to generate predictions for the hadronic 
parameters $\{m,\,\Gamma\}$. The following combinations of the moments~(\ref{moments})
(where we suppress the explicit dependence on $\tau$ and $s_0$) are useful in this 
regard:
\begin{equation}\label{moment_combs}
\begin{gathered}
  \frac{M_{k,1}}{M_{k,0}}\quad,\quad
  \sigma^2_k \equiv \frac{M_{k,2}}{M_{k,0}} -\left(\frac{M_{k,1}}{M_{k,0}}\right)^2 \\
  A_{k}^{(3)}\equiv \frac{M_{k,3}}{M_{k,0}} - 3\left(\frac{M_{k,2}}{M_{k,0}}\right)
  \left( \frac{M_{k,1}}{M_{k,0}}\right) + 2\left(\frac{M_{k,1}}{M_{k,0}}\right)^3
  \quad.
\end{gathered}
\end{equation}
In the spirit of the sum-rules approach,  we equate each of these theoretical
quantities to its corresponding phenomenological counterpart
resulting in the following system of equations:
\begin{gather}
  \frac{M_{0,1}}{M_{0,0}} = m^2 \label{square_params_a}\\
  \sigma_0^2-2\tau =\frac{1}{3}m^2 \Gamma^2 \label{square_params_b}\\
  A_0^{(3)} = 0 \label{square_params_c}\quad.
\end{gather}
Considering that the right-hand (phenomenological) 
sides of~(\ref{square_params_a})--(\ref{square_params_c}) are all $\tau$-independent,
it is an important feature of the analysis that, at the optimized continuum threshold $s_0$,
the left-hand (theoretical) sides exhibit negligible dependence on $\tau$
and so are well approximated throughout the interval $[\tau_{\text{i}},\tau_{\text{f}}]$
by averaged values. Such $\tau$-independence of appropriate residual moment
combinations also occurs in the more complicated models considered below.

Then, we can use~(\ref{square_params_a})--(\ref{square_params_b}) to obtain
predictions for the two free parameters $\{ m,\,\Gamma\}$ of the square pulse 
model~(\ref{square_pulse}). Equation~(\ref{square_params_c}) is independent 
of our parameter estimates and therefore serves as an \textit{a posteriori} 
consistency test.\footnote{For the square pulse model, it seems we could
use either~(\ref{square_params_a}) or~(\ref{single_mass}) to compute the mass $m$; 
however, as we shall see, both yield the same result. 
We concentrate on~(\ref{square_params_a}) simply because
the analysis techniques outlined here are easily generalized 
to the more complicated resonance models to follow.}

Inverting~(\ref{square_params_a})--(\ref{square_params_b}), we obtain 
$m=(1.30\pm 0.17)\ \mbox{GeV}$ and $\Gamma=(0.59\pm 0.07)\ \mbox{GeV}$, 
where, in this and in all subsequent analyses, the uncertainties quoted stem from the 
error bounds on the dimension-four gluonic condensate~(\ref{dimfour})
and an estimated 15\% uncertainty in each of $n_{\text{c}}$ and $\rho_{\text{c}}$ 
(see~(\ref{DILparams})).
In Figure~\ref{square_fig}, we use $s_0=2.3\ \mbox{GeV}^2$ and 
central values of $m$ and $\Gamma$ to plot the theoretical and 
phenomenological sides of~(\ref{gauss_sp}) versus $\hat{s}$ for 
$\tau\in\{ 2,3,4 \}\ \text{GeV}^4$.
The excellent agreement between theory and phenomenology demonstrated 
by these plots is a vast improvement over the results of the $k=0$ 
single narrow resonance analysis (see Figure~\ref{sum_rules_single}).

However, closer quantitative scrutiny reveals that the fits depicted in 
Figure~\ref{square_fig} are not quite as accurate as one might think. 
The quantity $A_0^{(3)}$ involves higher order moments of the $k=0$ 
Gaussian sum-rule and so is sensitive to an even finer level of detail 
than are $M_{0,1}/M_{0,0}$ and $\sigma_0^2$. A QCD calculation of this
quantity yields $A_0^{(3)}=-0.0825\,{\rm GeV^6}$ in significant violation of our
consistency check~(\ref{square_params_c}).

For our second example of a single non-zero width model, we would ideally like 
to consider a Breit-Wigner resonance. However, closed-form expressions do not exist 
for the Gaussian image of a Breit-Wigner shape, and so we instead employ a Gaussian 
model to describe a well-defined resonance peak with a non-zero width
\begin{equation}
 \rho^{\text{had}}(t)= \rho^{\text{g}}(t)\equiv 
 f^2_g\exp{\left[-\frac{\left(t-m^2\right)^2}{2\Gamma^2}\right]}\quad .
\label{gauss_res}
\end{equation}
The quantity $f_g^2$ is a normalization constant related to the total resonance strength, 
and the Gaussian width $\Gamma$ 
is related to an equivalent Breit-Wigner width $\Gamma_{BW}$ through
\begin{equation}
\Gamma_{BW}=\sqrt{2\log 2}\frac{\Gamma}{m}\quad.
\label{Gamma_bw}
\end{equation}
In this model, the $k=0$ normalized Gaussian sum-rule~(\ref{ngsr}) becomes
\begin{equation}
  N^{\qcd}_0(\hat{s},\tau,s_0) =
  \frac{1+\text{erf}\left(\frac{\hat s\Gamma^2+2m^2\tau}{2\Gamma\sqrt{\tau}\sqrt{\Gamma^2+2\tau}}\right)}
  {\sqrt{2\pi}\sqrt{\Gamma^2+2\tau}\left[1+\text{erf}\left(\frac{m^2}{\sqrt{2}{\Gamma}}\right)\right]}
  \exp\left[-\frac{\left(\hat{s}-m^2\right)^2}{2\left(\Gamma^2+2\tau\right)}\right]
\quad ,
\label{phenom_gauss}
\end{equation}
where 
\begin{equation}
{\rm erf}(x)=\frac{2}{\sqrt{\pi}}\int\limits_0^x e^{-y^2}\,dy\quad .
\label{erf}
\end{equation}

As in prior analyses, we begin with an extraction of the optimized continuum 
threshold parameter $s_0$, and again, we accomplish this by examining
the behaviour or the peak position. The phenomenological side of~(\ref{phenom_gauss})
admits a single peak; however, in contrast to our previous two analyses,
\textit{the position of this peak is dependent on $\tau$.}
Differentiating the right-hand side of~(\ref{phenom_gauss}) and setting the result to
zero gives
\begin{equation}
  \hat s-m^2=\frac{\Gamma}{\sqrt{\pi\tau}}\left[
  \frac{\exp\left(-\frac{\left(\hat s\Gamma^2+2m^2\tau\right)^2}{4\Gamma^2\tau
  \left(\Gamma^2+2\tau\right)}\right)}{1+\text{erf}\left(\frac{\hat 
  s\Gamma^2+2m^2\tau}{2\Gamma\sqrt{\tau}\sqrt{\Gamma^2+2\tau}}\right)}
  \right]
\label{gauss_constraint}
\end{equation}
which, unfortunately cannot be explicitly solved for $\hat{s}$.
Consequently, in the absence of an exact solution, 
we approximate the phenomenological peak position by the expression
\begin{equation}\label{peak_approx}
  A + \frac{B}{\tau} + \frac{C}{\tau^2}
\end{equation} 
where $\{A,B,C\}$ are to be considered unknown   parameters.
Explicit numerical experiments in (realistic) worst-case scenarios show that,
provided $\tau\geq 2\ \unit{GeV}^4$,  the next term
in the expansion~(\ref{peak_approx}) [\emph{i.e.}\ $D/\tau^3$] is negligible
and can safely be ignored.
Therefore, we are led to define the following
$\chi^2$-function as a measure of the deviation of the theoretical peak position~(\ref{peakpos}) 
from the phenomenological peak position characterized by the expansion~(\ref{peak_approx}):
\begin{equation}\label{double_chi}
  \chi^2(s_0,A,B,C) = \sum_{n=0}^N \left[ \hat{s}_{\text{peak}}(\tau_n, s_0) 
  - A - \frac{B}{\tau_n} - \frac{C}{\tau_n^2} \right]^2 \quad.
\end{equation}
The $\chi^2$ minimizing conditions
\begin{equation}\label{min}
  \frac{\partial \chi^2}{\partial A} = \frac{\partial \chi^2}{\partial B} = 
  \frac{\partial \chi^2}{\partial C} = 0
\end{equation} 
can then  be used to write $\{A,B,C\}$ as functions\footnote{The set of equations defined
by condition~(\ref{min}) is linear and inhomogeneous. While trivial to obtain, the solution
is rather a mess and so is omitted for brevity.} of $s_0$
leaving us with a one-dimensional minimization problem in $s_0$:
\begin{equation}\label{oneD}
  \chi^2(s_0) = \sum_{n=0}^N \left[ \hat{s}_{\text{peak}}(\tau_n, s_0)
  - A(s_0) - \frac{B(s_0)}{\tau_n} - \frac{C(s_0)}{\tau_n^2} \right]^2 \quad.
\end{equation}
Minimizing~(\ref{oneD}) with respect to $s_0$ furnishes us with an optimized choice for 
the continuum threshold parameter. 

Once the optimized $s_0$ has been determined, the subsequent analysis of the Gaussian resonance
model proceeds in a fashion completely analogous to the analysis of the square pulse model. 
For instance, the appropriate Gaussian model equations corresponding 
to~(\ref{square_params_a})--(\ref{square_params_c}) are
\begin{gather}
  \frac{M_{0,1}}{M_{0,0}} =m^2+\Gamma\Delta\label{gauss_m1} \\
  \sigma_0^2-2\tau=\Gamma^2-m^2\Gamma\Delta-\Gamma^2\Delta^2 \label{gauss_sigma} \\
  A^{(3)}_0=\left(m^4\Gamma-\Gamma^3\right)\Delta+3m^2\Gamma^2\Delta^2+2\Gamma^3\Delta^3 \label{gauss_A}
\end{gather}
where
\begin{equation}\label{deldef}
  \Delta=\sqrt{\frac{2}{\pi}}\left[
  \frac{\exp{\left(-\frac{m^4}{2\Gamma^2}\right)}}{1+\text{erf}
  \left(\frac{m^2}{\sqrt{2}\Gamma} \right)}\right]\quad.
\end{equation}
Again, the first two equations may be (numerically) inverted to yield predictions for the 
model's two free parameters $\{ m,\,\Gamma\}$  while the third serves as an independent
consistency check.  The quantity $\Delta$ is small, and hence the terms
in~(\ref{gauss_m1}--\ref{gauss_A}) proportional to $\Delta$ have a
negligible effect on the extracted resonance parameters, a property which
also simplifies our subsequent analysis.  Note that since $A_0^{(3)}\sim
\Delta$ in (\ref{gauss_A}), we anticipate that this model will still
underestimate the QCD value of this moment combination, a result which is
confirmed  in our detailed analysis.

Carrying out the appropriate sequence of calculations first gives an optimized continuum 
threshold parameter\footnote{Surprisingly, this is the same optimized threshold parameter
value that we found in the single narrow resonance model analysis. This need not always 
be the case as depicted explicitly in~\cite{orl00}.} 
$s_0=2.3\ \mbox{GeV}^2$ which, in turn, yields a mass of $m=(1.30\pm 0.17)\ \mbox{GeV}$ 
and an equivalent Breit-Wigner width (see~(\ref{Gamma_bw}))
of $\Gamma_{\text{BW}}=(0.40\pm 0.05)\ \mbox{GeV}$.
Again, we test the validity of these
values by plotting both the theoretical and phenomenological sides 
of~(\ref{phenom_gauss}) versus $\hat{s}$ for $\tau\in\{ 2,3,4\}\ \text{GeV}^4$, 
setting $s_0=2.3\ \mbox{GeV}^2$ and employing the central values of 
the resonance parameters.
The resulting graphs are, to the naked eye, indistinguishable from those corresponding to 
the square pulse model analysis\footnote{We quantitatively address this
situation in Section~\ref{discussion}.} (Figure~\ref{square_fig}) and so are omitted.

Substitution of the central values of $m$ and $\Gamma$ into~(\ref{gauss_A}) 
yields $A_0^{(3)}=0.000342\,{\rm GeV^6}$ which must again be compared with the QCD value
of $A_0^{(3)}=-0.0825\,{\rm GeV^6}$. Clearly, there exists a significant discrepancy between the 
two as the Gaussian model cannot even correctly predict the sign of this 
moment combination.

The quantity $A_0^{(3)}$, however, is a measure of the asymmetry of 
$N_0^{\text{QCD}} (\hat{s},\tau,s_0)$ with respect to $\hat{s}$ 
about its average value defined by $M_{0,1}/M_{0,0}$.  Therefore,
considering that both the Gaussian and square pulse models represent
resonance strength distributions that are symmetric (about $m^2$), it is perhaps
not surprising that they fail to accurately predict $A_0^{(3)}$.  
Hence, we are prompted to consider a skewed generalization of 
the Gaussian resonance model:
\begin{equation}\label{skewed_gauss}
  \rho^{\text{had}}(t) = \rho^{\text{sg}}(t) 
  \equiv t^2 f^2 \exp\left[ -\frac{(t-m^2)^2}{2\Gamma^2} \right]
\end{equation}
where the factor $t^2$ introduces a degree of asymmetry
and has been chosen to achieve consistency with known low-energy two pion 
decay rates~\cite{let,pion}.
Substituting~(\ref{skewed_gauss}) into~(\ref{ngsr}) gives
\begin{equation}\label{sr_skewed_gauss}
\begin{aligned}
  E\,N_0^{\qcd}(\hat{s},\tau,s_0) &= 
  2\sqrt{\pi}\exp\left[-\frac{(\hat{s}-m^2)^2}{2(\Gamma^2 +2\tau)}\right]
  (\hat{s}^2\Gamma^4 +4\hat{s}\Gamma^2m^2\tau +4m^4\tau^2 +2\tau\Gamma^4 +4\tau^2\Gamma^2)
  \left[ 1+\erf\left(\frac{\hat{s}\Gamma^2 
  +2\tau m^2}{2\Gamma\sqrt{\tau}\sqrt{\Gamma^2 +2\tau}}\right)\right] \\
  &+4\exp\left[-\frac{\hat{s}^2\Gamma^2 +2\tau m^4}{4\tau\Gamma^2}\right]
  \Gamma\sqrt{\tau}\sqrt{\Gamma^2 +2\tau}(\hat{s}\Gamma^2+2\tau m^2)
\end{aligned}
\end{equation}
where
\begin{equation}\label{skewed_gauss_denom}
  E=\sqrt{2\pi}(m^4+\Gamma^2)\left[1+\erf\left(\frac{m^2}{\sqrt{2}\Gamma}\right)\right]
  + 2m^2\Gamma\exp\left(-\frac{m^4}{2\Gamma^2}\right)\sqrt{\pi}(\Gamma^2
  +2\tau)^{5/2}
\quad .
\end{equation}
The sum-rules analysis proceeds along the same lines as for the 
unskewed Gaussian resonance model. The appropriate generalizations
of~(\ref{gauss_m1})--(\ref{gauss_A}) are
\begin{gather}
  \frac{M_{0,1}}{M_{0,0}} = \frac{m^2(m^4+3\Gamma^2)}{m^4+\Gamma^2}
  + \mathcal{O}(\Delta) \label{skew_m1}\\
  \sigma_0^2-2\tau = \frac{\Gamma^2(m^8+3\Gamma^4)}{(m^4+\Gamma^2)^2}
  + \mathcal{O}(\Delta) \label{skew_sigma}\\
  A_0^{(3)} = \frac{4m^2\Gamma^6(m^4-3\Gamma^2)}{(m^4+\Gamma^2)^3}
  + \mathcal{O}(\Delta) \label{skew_A}
\end{gather}
Numerically inverting\footnote{The terms in
  ~(\protect\ref{skew_m1})--(\protect\ref{skew_sigma}) proportional to $\Delta$ are found to
  have a negligible impact on the solutions.  }
~(\ref{skew_m1})--(\ref{skew_sigma}), we find a mass $m=(1.17\pm 0.15)\,{\rm GeV}$
and an equivalent Breit-Wigner width of $\Gamma_{\text{BW}}=(0.49\pm 0.06)\,{\rm GeV}$. 
A subsequent phenomenological prediction  yields 
$A_0^{(3)}=0.00943\,{\rm GeV^6}$---again, completely inaccurate compared with
  the QCD value.
These results, as well as the results of the unskewed non-zero resonance width model
analyses are summarized for convenience in Table~\ref{width_tab}.  
\begin{table}[htb]
  \centering
 
  \begin{tabular}{||c|r|r|r||}
    \hline\hline
    resonance model & mass (GeV) & width (GeV) & $A^{(3)}_0~({\rm GeV^6})$ \\ 
    \hline\hline
    unskewed Gaussian & 1.30 & 0.38 & 0.000342 \\\hline
    unskewed square pulse & 1.30 & 0.59 & 0 \\\hline
    skewed Gaussian & 1.18 & 0.49 & 0.00943 \\
    \hline\hline
  \end{tabular}
 \caption{The results of a $k=0$ Gaussian sum-rules analysis
           of a variety of non-zero  resonance width models using central values of the QCD parameters.
    For the Gaussian resonance models,
    the given width is actually the equivalent Breit-Wigner width 
    (see~(\protect{\ref{Gamma_bw}})).  The $A_0^{(3)}$ values should be compared with the QCD prediction 
$A_0^{(3)}=-0.0825\,{\rm GeV^6}$.
}
\label{width_tab}
\end{table}

\subsubsection{Double Narrow Resonance Model}\label{dnr}
The double narrow resonance model is defined by
\begin{equation}\label{double}
  \rho^{\text{had}}(t) = \rho^{\text{2r}}(t)\equiv\pi 
  \left[f_1^2 \delta(t-m_1^2) + f_2^2 \delta(t-m_2^2) \right]
\end{equation}
where $m_1\leq m_2$ are the two resonance masses and $f_1,f_2$ are their respective couplings.
Correspondingly, the normalized Gaussian sum-rule~(\ref{ngsr}) reduces to 
\begin{equation} \label{phenom_double}
  N^{\qcd}_k(\hat{s},\tau,s_0)  = \frac{1}{\sqrt{4\ppi\tau}}
  \left\{ \frac{f_1^2 m_1^{2k}}{f_1^2 m_1^{2k}+f_2^2 m_2^{2k}} \exp\left[-\frac{(\hat{s}-m_1^2)^2}{4\tau}\right]
  +       \frac{f_2^2 m_2^{2k}}{f_1^2 m_1^{2k}+f_2^2 m_2^{2k}} \exp\left[-\frac{(\hat{s}-m_2^2)^2}{4\tau}\right]
  \right\}\quad.
\end{equation}
In the analysis of this model, it becomes inconvenient
to use the parameters $\{m_1,f_1,m_2,f_2\}$ and so we instead focus on the set $\{z,y,r\}$ defined by
\begin{equation}\label{new_params}
  z  =  m_1^2 + m_2^2 \quad,\quad
  y  =  m_1^2 - m_2^2 \quad,\quad
  r  =  r_1 - r_2
\end{equation}
with
\begin{equation}\label{crud}
  r_1 =  \frac{f_1^2 m_1^{2k}}{f_1^2 m_1^{2k}+f_2^2 m_2^{2k}} \quad,\quad
  r_2 =   \frac{f_2^2 m_2^{2k}}{f_1^2 m_1^{2k}+f_2^2 m_2^{2k}}\quad\text{with}\quad
  r_1+r_2=1\quad.
\end{equation}

As in the Gaussian resonance model analysis, the phenomenological side of~(\ref{phenom_double}) 
admits a single peak whose position is $\tau$-dependent. 
In terms of the double resonance model parameters~(\ref{new_params}),
differentiating the right-hand side of~(\ref{phenom_double}) with respect to $\hat{s}$ 
and setting the result to zero yields
\begin{equation}
   \frac{(r+1)\left(\hat s-\frac{1}{2}z-\frac{1}{2}y\right)}{(r-1)\left(\hat s-\frac{1}{2}z
  +\frac{1}{2}y\right)}
  - \exp\left[ \frac{y\left(z-2\hat s\right)}{4\tau} \right] = 0 
\end{equation}
which, again, cannot be explicitly solved for $\hat{s}$. 
Thus, we approximate the phenomenological peak position by the 
expression~(\ref{peak_approx}) and correspondingly, extract an optimized 
continuum threshold parameter $s_0=2.3\ \text{GeV}^2$.

To compute predictions for the hadronic parameters of the double narrow resonances
model, we again look to moment combinations such as~(\ref{moment_combs}). 
However, since this model contains three free parameters~(\ref{new_params}), 
we require three equations.
In other words, if we wish to have an independent consistency check of our results, 
we must introduce a fourth moment combination in addition to those of~(\ref{moment_combs}).
Defining
\begin{equation}\label{skmom}
  S_k\equiv \frac{M_{k,4}}{M_{k,0}}-4\frac{M_{k,3}}{M_{k,0}}\frac{M_{k,1}}{M_{k,0}}
  +6\frac{M_{k,2}}{M_{k,0}}\left(\frac{M_{k,1}}{M_{k,0}}\right)^2-3
  \left( \frac{M_{k,1}}{M_{k,0}}\right)^4
\end{equation}
leads to the following equations 
\begin{gather}
  \frac{M_{k,1}}{M_{k,0}} = \frac{1}{2}(z+ry)\label{double_m1}\\
  \sigma_k^2-2\tau = \frac{1}{4}y^2(1-r^2) \label{double_sigma}\\
  A_k^{(3)} = -\frac{1}{4}ry^3(1-r^2)\label{double_A}\\
  S_k-12\tau^2 -12\tau\left(\sigma_k^2 -2\tau\right) 
  =\frac{1}{16}y^4\left(1-r^2\right)\left(1+3 r^2\right)\label{doule_s} \quad.
\end{gather}
which are analogous to~(\ref{square_params_a})--(\ref{square_params_c})  
and~(\ref{gauss_m1})--(\ref{gauss_A}).

For the $k=0$ case of interest, upon inverting~(\ref{double_m1})--(\ref{double_A})
we find the heavier of the two states is also the more strongly coupled with 
$m_2= (1.4\pm 0.2)\ \unit{GeV}$ and $r_2=0.72\pm 0.06$.  
The lighter resonance with a mass $m_1=\left(0.98\pm 0.2\right)\,{\rm
  GeV}$  is the more weakly coupled
state with $r_1=0.28\mp 0.06$.
The QCD uncertainties given for the resonance masses obscure the very stable
  mass splitting between the two states: $m_2-m_1=(0.42\pm 0.03)\ \unit{GeV}$. 
Again, plots of the theoretical and phenomenological
Gaussian sum-rules exhibit the excellent agreement shown in Figure~\ref{square_fig}.

Using central values of the hadronic parameters obtained, we can calculate 
a phenomenological prediction associated  with the first independent moment 
combination~(\ref{skmom}). The result is
$S_0-12\tau^2-12\tau\left(\sigma_0^2-2\tau\right)= 0.073733\,{\rm GeV^8}$ 
which must compared with a QCD calculation of 
$S_0-12\tau^2-12\tau\left(\sigma_0^2-2\tau\right)= 0.169769\,{\rm GeV^8}$.
The  roughly 50\% deviation between these values  is a large improvement 
over similar comparisons in the single non-zero width resonance models.

\subsubsection{Two Resonance Model of a  Narrow Resonance Combined with a Wide Resonance}
A natural extension of the double narrow resonance model is to introduce 
an additional parameter into the hadronic model which describes the width for one of the
resonances.

We begin with a model consisting of a narrow resonance of mass $m$
and a Gaussian resonance of mass $M$ and width $\Gamma$.  The resulting $k=0$
normalized Gaussian sum-rule~(\ref{ngsr}) is:
\begin{equation}  
N_0^{\qcd}\left(\hat s, \tau,s_0\right)=
r_m\frac{1}{\sqrt{4\ppi\tau}}
 \exp\left[-\frac{(\hat{s}-m^2)^2}{4\tau}\right]
+r_M
\exp\left(-\frac{\left(\hat s -M^2\right)^2}{2\left(\Gamma^2+2\tau\right)}\right)
\left[
\frac{1+\text{erf}\left(\frac{\hat s\Gamma^2+2M^2\tau}{2\Gamma\sqrt{\tau}\sqrt{\Gamma^2+2\tau}}\right)}
{\sqrt{2\pi}\sqrt{\Gamma^2+2\tau}\left[1+\text{erf}\left(\frac{M^2}{\sqrt{2}{\Gamma}}\right)\right]}
\right] \quad ,
\label{N_narrow_gauss}
\end{equation}
where $r_m$ and $r_M$ denote the relative strengths of the resonances and are
constrained by $r_m+r_M=1$.  As in the other models, the phenomenological side
of~(\ref{N_narrow_gauss}) has a single $\tau$-dependent peak position. The peak-drift $\chi^2$
in~(\ref{double_chi}) is again minimized to find the optimum value of the continuum,
and then the theoretical (QCD) values for the moments are
used to determine the resonance parameters.  Since this model contains four
independent parameters, the lowest four moment combinations~(\ref{moment_combs},\ref{skmom}) are used to
determine the parameters and the following  fifth order moment combination
will be used as a consistency check. 
\begin{equation}
A_k^{(5)}= \frac{M_{k,5}}{M_{k,0}}-5 \frac{M_{k,4}}{M_{k,0}}
\frac{M_{k,1}}{M_{k,0}}
+10 \frac{M_{k,3}}{M_{k,0}}\left( \frac{M_{k,1}}{M_{k,0}}\right)^2
-10 \frac{M_{k,2}}{M_{k,0}}\left( \frac{M_{k,1}}{M_{k,0}}\right)^3
+4\left( \frac{M_{k,1}}{M_{k,0}} \right)^5
\label{fifth_moment}
\end{equation}
 Defining
\begin{equation}
z=M^2+m^2\quad ,\quad y=m^2-M^2\quad ,\quad r=r_m-r_M
\label{narrow_gauss_defs}
\end{equation}
we find the following expressions for the moment combinations in terms of the resonance
parameters
\begin{gather}
 \frac{M_{0,1}}{M_{0,0}}=\frac{1}{2}\left( z+ ry \right)+ {\cal O}\left(\Delta\right)
\label{narrow_gauss_m1}
\\
\sigma_0^2-2\tau=\frac{1}{4}y^2\left(1-r^2\right)+\frac{1}{2}\Gamma^2\left(1-r\right)
+ {\cal O}\left(\Delta\right)
\label{narrow_gauss_m2}
\\
A_0^{(3)}=-\frac{1}{4}y^3r\left(1-r^2\right)-\frac{3}{4}\Gamma^2
y\left(1-r^2\right)
+ {\cal O}\left(\Delta\right)
\label{narrow_gauss_m3}
\\
S_0-12\tau^2-12\tau\left(\sigma_0^2-2\tau\right)=\frac{1}{16}y^4\left(1-r^2\right)\left(1+3r^2\right)
+\frac{3}{4}\Gamma^2y^2\left(1+r\right)\left(1-r^2\right)
+\frac{3}{2}\Gamma^4\left(1-r\right)
+ {\cal O}\left(\Delta\right)
\label{narrow_gauss_m4}
\\
A_0^{(5)}-20\tau A_0^{(3)}=-\frac{1}{8}y^5r\left(1-r^2\right)\left(1+r^2\right)
-\frac{5}{8}\Gamma^2y^3\left(1-r\right)\left(1+r\right)^3
-\frac{15}{4}\Gamma^4y\left(1-r^2\right)
+ {\cal O}\left(\Delta\right)
\label{narrow_gauss_m5}
\end{gather}
The terms proportional to $\Delta$ (see equation~(\ref{deldef})) in the above expressions are found to be
numerically insignificant.

The QCD  values of the moments are used to 
solve~(\ref{narrow_gauss_m1}--\ref{narrow_gauss_m4}) for the resonance
parameters.\footnote{As mentioned earlier, the QCD value of the residual
moment combinations are found to be virtually $\tau$ independent consistent
with the resonance model expressions~(\protect\ref{narrow_gauss_m1}--\protect\ref{narrow_gauss_m5}).}  
The  non-linear nature of~(\ref{narrow_gauss_m1}--\ref{narrow_gauss_m4})
might suggest a large number of solutions, but it is found that only one
physical solution persists across the range of QCD parameters considered:
$m=(1.41\pm 0.19)\,{\rm GeV}$, $M=(1.23\pm 0.15)\,{\rm GeV}$, 
$\Gamma_{BW}=(0.52\pm 0.06)\,{\rm GeV}$, $r_m=0.49\pm 0.13$, $r_M=1-r_m$. 
Plots of  the theoretical and phenomenological sides of the normalized
Gaussian sum-rules are again represented in Figure~\ref{square_fig}, illustrating excellent agreement
between the QCD prediction and phenomenological model.

The fifth-order moment~(\ref{narrow_gauss_m5}) can now be used as a
consistency check.  The resonance parameters for the central QCD parameters
yields $A_0^{(5)}-20\tau A_0^{(3)}=-0.130\, {\rm GeV^{10}}$ which should be compared with the
QCD value of $A_0^{(5)}-20\tau A_0^{(3)}=-0.243 {\rm GeV^{10}}$.

To assess whether other non-zero width models lead to better agreement with
the QCD value of $A_0^{(5)}-20\tau A_0^{(3)}=-0.243\, {\rm GeV^{10}}$ we consider using the square
pulse model instead of the Gaussian resonance, which results in a normalized
Gaussian sum-rule similar to~(\ref{N_narrow_gauss})
\begin{equation}
 N^{\qcd}_0(\hat{s},\tau,s_0)
=r_m
\frac{1}{\sqrt{4\ppi\tau}}
 \exp\left[-\frac{(\hat{s}-m^2)^2}{4\tau}\right]
+r_M\frac{1}{4M\Gamma}\left[
{\rm erf}\left(\frac{\hat s-M^2+M\Gamma}{2\sqrt{\tau}}\right)
-{\rm erf}\left(\frac{\hat s-M^2-M\Gamma}{2\sqrt{\tau}}
\right)
\right]\quad ,
\label{N_narrow_square}
\end{equation}
where $r_m+r_M=1$. Using the definitions~(\ref{narrow_gauss_defs}) results in
the following expressions for the moment combinations in terms of the
resonance parameters.
\begin{gather}
 \frac{M_{0,1}}{M_{0,0}}=\frac{1}{2}\left( z+ ry \right)+ {\cal O}\left(\Delta\right)
\label{narrow_square_m1}
\\
\sigma_0^2-2\tau=\frac{1}{4}y^2\left(1-r^2\right)+\frac{1}{12}\Gamma^2\left(z-y\right)\left(1-r\right)
+ {\cal O}\left(\Delta\right)
\label{narrow_square_m2}
\\
A_0^{(3)}=-\frac{1}{4}y^3r\left(1-r^2\right)-\frac{1}{8}\Gamma^2
y\left(1-r^2\right)\left(z-y\right) + {\cal O}\left(\Delta\right)
\label{narrow_square_m3}
\\
\begin{split}
S_0-12\tau^2-12\tau\left(\sigma_0^2-2\tau\right)=&\frac{1}{16}y^4\left(1-r^2\right)\left(1+3r^2\right)
+\frac{1}{8}\Gamma^2y^2\left(1+r\right)\left(1-r^2\right)\left(z-y\right)
\\&
+\frac{1}{40}\Gamma^4\left(1-r\right)\left(z-y\right)^2+ {\cal
  O}\left(\Delta\right)
\end{split}
\label{narrow_square_m4}
\\
\begin{split}
A_0^{(5)}-20\tau A_0^{(3)}=&-\frac{1}{8}y^5r\left(1-r^2\right)\left(1+r^2\right)
-\frac{5}{48}\Gamma^2y^3\left(1-r^2\right)\left(1+r\right)^2\left(z-y\right)
\\
&-\frac{1}{16}\Gamma^4y\left(1-r^2\right)\left(z-y\right)^2+ {\cal O}\left(\Delta\right)
\end{split}
\label{narrow_square_m5}
\end{gather} 

Using the QCD values of the moments to solve~(\ref{narrow_square_m1}--\ref{narrow_square_m4}) 
for the resonance parameters again yields a single physical solution over the
QCD parameter space: 
$m=(1.33\pm 0.18)\,{\rm GeV}$, $M=(1.23\pm 0.18)\,{\rm GeV}$, 
$\Gamma_{BW}=(0.95\pm 0.12)\,{\rm GeV}$, $r_m=0.60\pm 0.13$, $r_M=1-r_m$. 
Agreement between the theoretical and phenomenological sides of the normalized
Gaussian sum-rules is again excellent, as exhibited by the plots shown in Figure~\ref{square_fig}.

The value of the fifth-order moment combination~(\ref{narrow_square_m5}) 
for  the resonance parameters corresponding to  the central QCD parameters
is $A_0^{(5)}-20\tau A_0^{(3)}=-0.114\, {\rm GeV^{10}}$. The square pulse model is thus not an
improvement upon the Gaussian resonance model's agreement with the QCD value
of this residual moment combination.

As a final scenario, we modify 
 the normalized
Gaussian sum-rule of~(\ref{N_narrow_gauss}) for a skewed Gaussian resonance
\begin{equation}
\begin{split}
 N^{\qcd}_0(\hat{s},\tau,s_0)
=&r_m
\frac{1}{\sqrt{4\ppi\tau}}
 \exp\left[-\frac{(\hat{s}-m^2)^2}{4\tau}\right]
\\
&+\frac{r_M}{E} 
2\sqrt{\pi}\exp\left[-\frac{(\hat{s}-M^2)^2}{2(\Gamma^2 +2\tau)}\right]
  (\hat{s}^2\Gamma^4 +4\hat{s}\Gamma^2M^2\tau +4M^4\tau^2 +2\tau\Gamma^4 +4\tau^2\Gamma^2)
  \left[ 1+\erf\left(\frac{\hat{s}\Gamma^2 
  +2\tau M^2}{2\Gamma\sqrt{\tau}\sqrt{\Gamma^2 +2\tau}}\right)\right] \\
  &+\frac{r_M}{E} 4\exp\left[-\frac{\hat{s}^2\Gamma^2 +2\tau M^4}{4\tau\Gamma^2}\right]
  \Gamma\sqrt{\tau}\sqrt{\Gamma^2 +2\tau}(\hat{s}\Gamma^2+2\tau M^2)
\end{split}
\label{N_narrow_skew}
\end{equation}
where $E$ (with appropriate substitution of the mass $M$) is defined in~(\ref{skewed_gauss_denom}).
Using the same conventions, we obtain  lengthy expressions for the moment combinations in terms
of the resonance parameters.
\begin{gather}
D \frac{M_{0,1}}{M_{0,0}}=\frac{1}{2}\left(z+r
 y\right)\left(z-y\right)^2
+\Gamma^2\left(-2y+4z+4ry-2rz \right)+ {\cal O}\left(\Delta\right)
\label{narrow_skew_m1}
\\
D=\left(z-y\right)^2+4\Gamma^2
\label{D_def}
\end{gather}
\begin{equation}
\begin{split}
D^2\left(\sigma_0^2-2\tau\right)=&\frac{1}{4}y^2\left(1-r^2\right)\left(z-y\right)^4
+\frac{1}{2}\Gamma^2(1-r)\left(z-y\right)^2
\left(z^2-4ryz-6yz+9y^2+8ry^2\right)
\\
&+4\Gamma^4\left(1-r^2\right)\left(z-2y\right)^2
+24\Gamma^6(1-r)+ {\cal O}\left(\Delta\right)
\end{split}
\label{narrow_skew_m2}
\end{equation}
\begin{equation}
\begin{split}
D^3A_0^{(3)}=&-\frac{1}{4}y^3r\left(1-r^2\right)\left(z-y\right)^6
-\frac{3}{4}\Gamma^2y\left(1-r^2\right)\left(z-y\right)^4
\left(z^2-4zry-2yz+8ry^2+y^2\right)
\\
&+3\Gamma^4\left(1-r^2\right)\left(z-y\right)^2(z-2y)\left(z^2-4zry-2yz+8ry^2+y^2\right)
\\
&+4\Gamma^6(1-r)
\left[y^3\left(-32r^2-41r-13\right)+z^34\left(r^2+r+1\right)+y^2z6\left(8r^2+11r+5\right)
+z^2y\left(-24r^2-33r-21\right)
\right]
\\
&+48\Gamma^8(1-r)\left(-6ry+3rz-2y-z\right)+ {\cal O}\left(\Delta\right)
\end{split}
\label{narrow_skew_m3}
\end{equation}
\begin{equation}
\begin{split}
D^4&\left[S_0-12\tau^2 -12\tau\left(\sigma_0^2-2\tau\right)\right]=
\frac{1}{16}y^4\left(1-r^2\right)\left(1+3r^2\right)\left(z-y\right)^8
\\
&+\frac{1}{4}\Gamma^2y^2\left(1-r^2\right)\left(z-y\right)^8
\left(3rz^2+3z^2-6zry-12zr^2y-10yz+24r^2y^2+11y^2+3ry^2  \right)
\\
&+\frac{3}{2}\Gamma^4\left(1-r\right)\left(z-y\right)^4\left[
z^4+y^4\left(48r^3+56r^2+32r+25 \right)
+yz^3\left(-4r^2-8r-8  \right)
+y^3z\left(-48r^3-68r^2-56r-40 \right)\right.
\\
&\phantom{+\frac{3}{2}\Gamma^4\left(1-r\right)\left(z-y\right)^4}\quad
\left.+y^2z^2\left(12r^3+28r^2+36r+26 \right)
\right]
\\
&+4\Gamma^6\left(1-r^2\right)\left(z-y\right)^2\left[
z^43(1+r)+y^4\left(96r^2+21r+61\right)
+yz^3\left(-12r^2-18r-30\right)
+y^3z\left(-144r^2-54r-126 \right)\right.
\\
&\phantom{+4\Gamma^6\left(1-r^2\right)\left(z-y\right)^2}\quad
\left.+y^2z^2\left( 72 r^2+48 r+96\right)
\right]
\\
&+16\Gamma^8\left(1-r\right)\left[
z^4\left( 3r^3+3r^2+9r+12\right)+y^4\left(48r^3+84r^2+80r+47\right)
+y^3z\left(-96r^3-186r^2-196r-118\right)\right.
\\
&\phantom{+16\Gamma^8\left(1-r\right)}\quad
\left.+z^3y\left(-24r^3-42r^2-60r-54\right)
+y^2z^2\left(72r^3+144r^2+168r+114\right)
\right]
\\
&+192\Gamma^{10}\left(1-r\right)\left[
y^2\left(12r^2+8r+8\right)+z^2\left(3r^2-2r+7\right)
-12yz(1+r)
\right]
\\
&+1920\Gamma^{12}\left(1-r\right)+ {\cal O}\left(\Delta\right)
\end{split}
\label{narrow_skew_m4}
\end{equation}
\begin{equation}
\begin{split}
D^5&\left[A_0^{(5)}-20\tau
  A_0^{(3)}\right]=-\frac{1}{8}y^5r\left(1-r^2\right)\left(1+r^2\right)
\left(z-y\right)^{10}
\\
&-\frac{5}{8}\Gamma^2\left(1-r^2\right)\left(z-y\right)^8
\left[z^2\left(1+r\right)^2+y^2\left(8r^3+r^2+10r+1 \right)
+yz\left(-4r^3-2r^2-8r-2\right)
\right]
\\
&-\frac{5}{4}\Gamma^4y\left(1-r^2\right)\left(z-y\right)^6\left[
3z^4+y^4\left(64r^3+12r^2+88r+15\right)
+z^3y\left(-6r^2-12r-18 \right)
\right.
\\
&\phantom{-\frac{5}{4}\Gamma^4y\left(1-r^2\right)\left(z-y\right)^6}\quad
\left.+y^3z\left(-64r^3-30r^2-124r-42 \right)+y^2z^2\left(16r^3+24r^2+64r+42\right)
\right]
\\
&+5\Gamma^6\left(1-r^2\right)\left(z-y\right)^4\left[
3z^5+y^5\left(-128r^3-30r^2-196r-44\right)
+y^4z\left(192r^3+84r^2+384 r+135 \right)
\right.
\\
&\phantom{+5\Gamma^6\left(1-r^2\right)\left(z-y\right)^4}\quad
 +z^4y\left(-6r^2-12r-24\right)+z^3y^2\left(16r^3+36r^2+96r+86\right)
\\
&\phantom{+5\Gamma^6\left(1-r^2\right)\left(z-y\right)^4}\quad
\left.+y^3z^2\left(-96r^3-84r^2-288r-156 \right)\right]
\\
&+40\Gamma^8\left(1-r\right)\left(z-y\right)^2\left[
y^5\left(-64r^4-90r^3-146r^2-153r-37\right)+
z^5\left(r^3+3r^2+3r+5 \right)\right.
\\
&\phantom{+40\Gamma^8\left(1-r\right)\left(z-y\right)^2}\quad
+y^4z\left(128r^4+201r^3+367r^2+399r+125\right)
+z^4y\left(-4r^4-12r^3-36r^2-47r-39 \right)
\\
&\phantom{+40\Gamma^8\left(1-r\right)\left(z-y\right)^2}\quad
+z^3y^2\left(32r^4+66r^3+162r^2+202r+114\right)
\\
&\phantom{+40\Gamma^8\left(1-r\right)\left(z-y\right)^2}\quad
\left.+y^3z^2\left(-96r^4-170r^3-354r^2-408r-168\right)
\right]
\\
&+32\Gamma^{10}(1-r)\left[
y^5\left(-128r^4-308r^3-508r^2-558r-194 \right)
+z^5\left(4r^4+4r^3+24r^2+59r-1\right)\right.
\\
&\phantom{+32\Gamma^{10}(1-r)}\quad
+z^4y\left(-40r^4-85r^3-195r^2-485r-155 \right)
+y^4z\left(320r^4+860r^3+1420r^2+1755r+695 \right)
\\
&\phantom{+32\Gamma^{10}(1-r)}\quad
\left.+z^3y^2\left(160r^4+430r^3+750r^2+1460r+620\right)
+y^3z^2\left( -320r^4-905r^3-1495r^2-2235r-965\right)
\right]
\\
&+960\Gamma^{12}(1-r)\left[
z^3\left(2r^3-2r^2+14r-6\right)
+y^3\left(-16r^3-16r^2-37r-13\right)\right.
\\
&\phantom{+960\Gamma^{12}(1-r)}\quad
\left. +z^2y\left(-12r^3+4r^2-57r-1 \right)
+y^2z\left(24r^3+8r^2+74r+18\right)\right]
\\
&+3840\Gamma^{14}(1-r)\left[
y(-10r-2)+z(5r-3) 
\right]+ {\cal O}\left(\Delta\right)
\end{split}
\label{narrow_skew_m5}
\end{equation} 

Using the QCD values of the moments to solve~(\ref{narrow_skew_m1}--\ref{narrow_skew_m4}) 
for the resonance parameters again yields a single, stable,  physical solution over the
QCD parameter space: 
$m=(1.38\pm 0.13)\,{\rm GeV}$, $M=(1.06\pm 0.21)\,{\rm GeV}$, 
$\Gamma_{BW}=(0.69\pm 0.07)\,{\rm GeV}$, $r_m=0.44\pm 0.04$, $r_M=1-r_m$. 
Excellent agreement between the theoretical and phenomenological sides of the normalized
Gaussian sum-rules is illustrated by the plots shown in Figure~\ref{square_fig}.

The value of the fifth-order moment~(\ref{narrow_skew_m5}) 
for  the resonance parameters corresponding to  the central QCD parameters
is $A_0^{(5)}-20\tau A_0^{(3)}=-0.192\, {\rm GeV^{10}}$, a result which
has only a 21\% deviation from the  QCD value  $A_0^{(5)}-20\tau A_0^{(3)}=-0.243\, {\rm
  GeV^{10}}$. The skewed Gaussian
model is thus in
reasonable agreement
 with the QCD value  of the fifth-order residual moment combination.

\section{Results and Discussion}\label{discussion}
The resonance parameters devolving from the central values of the QCD
parameters have already been summarized in Table~\ref{width_tab} for the single wide
resonance models, and Table~\ref{res_summary_table}
summarizes the results for the resonance parameters obtained in the  double
resonance models. We note that in the narrow plus wide resonance scenarios, no
assumption was made on which state would be wide, and hence it is interesting that the analysis 
consistently predicts that the lightest state has the non-zero width.
Similarly,
Tables~\ref{width_tab} and \ref{moms_summary_table}   summarize the QCD and predicted values of the
moments which serve as a consistency check in each of the models.  Based on
this criteria, the double resonance models have much better agreement with the
QCD prediction than the single wide resonance models. The narrow plus wide
resonance model is the most accurate in this respect, with only a 21\%
deviation from the QCD value of the fifth order asymmetry moment 
 $A_0^{(5)}-20\tau A_0^{(3)}$.

\begin{table}
\centering
\begin{tabular}{||c|c|c|c|c||}
\hline\hline
Resonance Model & $m\,({\rm GeV})$ &  $M\,({\rm GeV})$ & Width (GeV) & $r=r_m-r_M$\\\hline\hline
Double Narrow & $1.41$ & $1.00$ & $0$ & $0.42$ \\\hline
Narrow plus Square & $1.34$ & $1.24$ & $0.93$ & $0.18$\\\hline
Narrow plus Gaussian & $1.38$ & $1.23$ & $0.51$ & $-0.10$\\\hline
Narrow plus Skewed Gaussian & $1.40$ & $1.00$ & $0.68$ & $-0.16$\\\hline\hline   
\end{tabular}
\caption{Resonance parameters obtained from central values of the QCD parameters 
in the various two-resonance scenarios. The mass $M$ denotes the state
associated with the quoted width. The width parameter quoted for the Gaussian
models is the equivalent Breit-Wigner width.}
\label{res_summary_table}
\end{table}

\begin{table}
\centering
\begin{tabular}{||c|c|c||}
\hline\hline
   & $S_0-12\tau^2-12\tau\left(\sigma_0^2-2\tau\right)~~({\rm GeV^8})$  &
   $A_0^{(5)}-20\tau A_0^{(3)}~~({\rm GeV^{10}})$  \\\hline\hline 
QCD & $0.1698$ & $-0.2433$\\\hline
Double Narrow & $0.0737$ & $-0.0502$\\
Narrow plus Square & ---& $-0.1144$ \\
Narrow plus Gaussian & ---& $-0.1300$\\
Narrow plus Skewed Gaussian &---& $-0.1918$ \\\hline\hline   
\end{tabular}
\caption{Comparison of the next-highest moment combinations  
in the  two-resonance scenarios with the QCD values. All values are obtained
from the central values of the QCD parameters.}
\label{moms_summary_table}
\end{table}

As illustrated in Figure~\ref{sum_rules_single}, the single narrow resonance model leads to a significant
deviation between the theoretical (QCD) and phenomenological results for the
normalized Gaussian sum-rule.  The  single wide resonance and double resonance models dramatically
improve this agreement, and all the models lead to plots which are
represented by Figure~\ref{square_fig}.     
However,  a $\chi^2$
measuring the difference between the theoretical and phenomenological curves
over the ranges of  $\hat s$ and $\tau$ used in
Figure~\ref{square_fig} can be used as another criteria to evaluate the
effectiveness of the phenomenological models since it  provides a quantitative measure of the difference between the theoretically 
and phenomenologically determined $k=0$ Gaussian sum-rules.
Table~\ref{chi_table} lists this $\chi^2$  at the optimized
value $s_0=2.3\ \mbox{GeV}^2$ in each of the resonance models considered for
the central QCD parameters.
We see that the double  resonance models provide a fit which is an order 
of magnitude better than any of the non-zero width resonance models. 
However, the $\chi^2$ is on the order of $10^{-6}$ for all the double narrow
resonance models, and hence does not provide a strong distinction between the
various scenarios.  

\begin{table}[htb]
\centering
\begin{tabular}{||c|c||}
  \hline\hline
  resonance model & $\chi^2$ \\
  \hline \hline
  unskewed Gaussian  & $1.21\times10^{-5}$ \\\hline
  unskewed square pulse  & $1.31\times10^{-5}$ \\\hline
  skewed Gaussian  & $1.45\times10^{-5}$ \\\hline
 double narrow  & $1.83\times10^{-6}$ \\\hline
narrow and square pulse & $0.923\times 10^{-6}$\\\hline
narrow and unskewed Gaussian & $0.888\times 10^{-6}$\\\hline 
narrow and skewed Gaussian & $0.959\times 10^{-6}$\\
  \hline\hline
\end{tabular}
\caption{\label{chi_table} The  $\chi$ for the fits between the theoretical and phenomenological
results for the normalized $k=0$ Gaussian sum-rules over the  
ranges of $\hat s$ and $\tau$ used in Figure \ref{square_fig} for
          each of the phenomenological models  considered. Central values of
          the QCD parameters have been employed.}
\end{table}

Thus the  moment consistency test and $\chi^2$ measure of the  agreement
between the phenomenological and QCD values of the Gaussian sum-rules clearly favour the double
resonance scenarios. However,  this $\chi^2$ does not clearly distinguish between the
various two-resonance scenarios.
Despite this apparent difficulty in distinguishing between the scenarios, 
the mass of the two states in the various double resonance scenarios
 are relatively stable, lying in the range $1.0\mbox{--}1.4\,{\rm GeV}$, with the
 greatest mass splittings occurring in the double narrow resonance model and in
 the narrow plus skewed Gaussian resonance model.

\section{Conclusions}\label{conclusion}
In this paper, we used QCD Gaussian sum-rules to analyze the scalar glueball 
sector in an effort to obtain predictions for the hadronic parameters 
({\it i.e.} mass, width, coupling strength) of low-lying scalar glueball states.
In our analysis, we incorporated instanton effects and employed a number of 
phenomenological models more general than the traditional single narrow 
resonance. 

First, we demonstrated that the leading order $k=-1$ Gaussian sum-rule 
(the only sum-rule which receives a contribution from the LET) 
leads to results which are unstable under moderate QCD uncertainties:
single resonance mass extractions ranged anywhere from 1.0 GeV to
1.8 GeV. This variation is too large to yield a definitive prediction
for the mass of the lightest scalar glueball, but does, however, indicate 
a lower bound of roughly 1 GeV on gluonium mass scales. 
This observation supports similar results obtained previously 
in~\cite{shuryak_forkel,har00} where it was shown that instanton effects
serve to increase the scale of masses extracted from the $k=-1$ 
sum-rule whereby reducing the discrepancy between masses predicted using
this $k=-1$ sum-rule and those extracted from $k\geq 0$ 
sum-rules.

\begin{figure}[htb]
\centering
\includegraphics[scale=0.65]{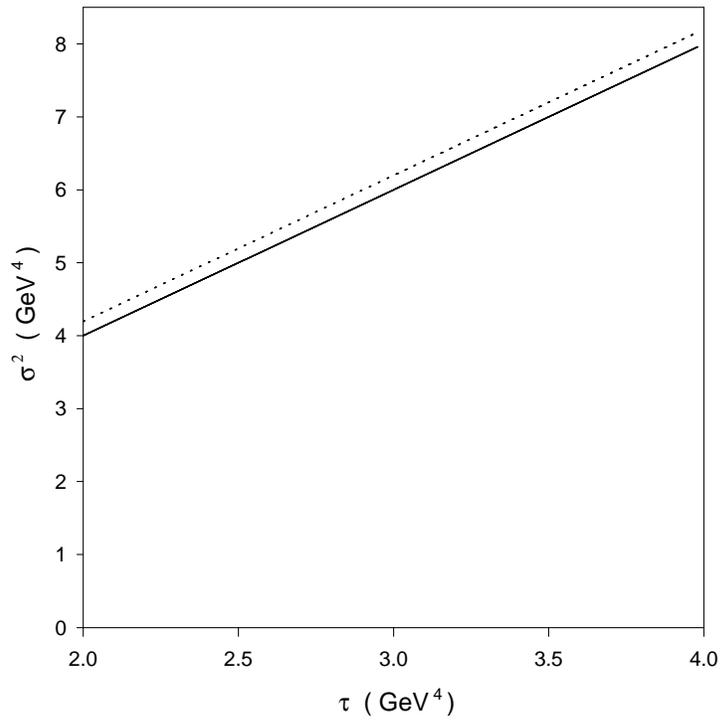}
\caption{Plot of $\sigma_0^2$ for the theoretical prediction (dotted curve) compared with $\sigma_0^2=2\tau$
for the single-resonance model (solid curve) for the $k=0$ sum-rule
using the $\chi^2$-optimized value of the continuum $s_0=2.3\,{\rm GeV^2}$.
\label{sigma}}
\end{figure}

\begin{figure}[htb]
\centering
\includegraphics[scale=0.65]{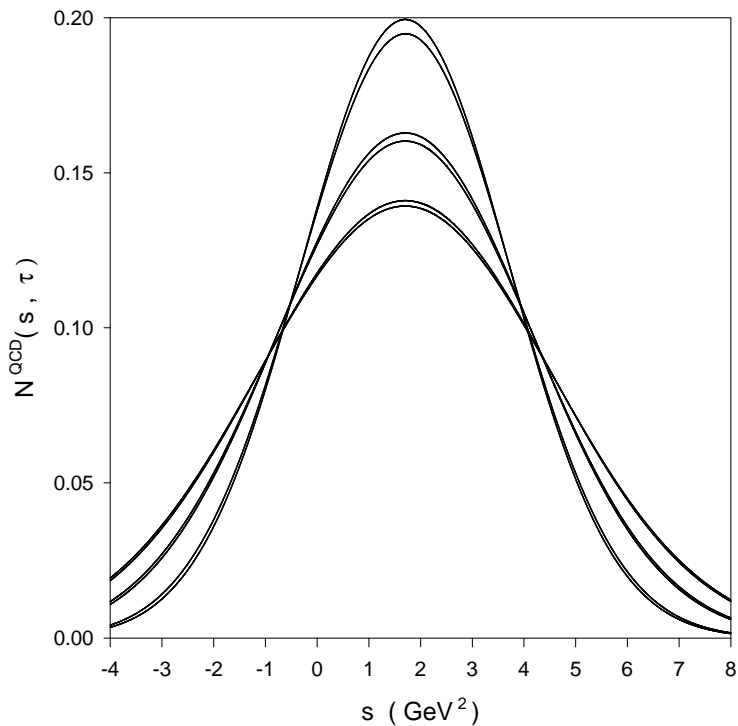}
\caption{Comparison of the
theoretical prediction for $N_0^{QCD}\left(\hat s, \tau,s_0\right)$ with the
single narrow resonance phenomenological model~(\protect\ref{phenom_single}) using
the $\chi^2$-optimized values of the resonance mass ($m=1.30\,{\rm GeV}$)
and continuum ($s_0=2.3\,{\rm GeV^2}$).
The $\tau$ values used for the three pairs of curves, from top to bottom in the figure, are respectively
 $\tau=2.0\ \unit{GeV}^4$, $\tau=3.0\ \unit{GeV}^4$, and $\tau=4.0\ \unit{GeV}^4$.
The phenomenological model is consistently {\em larger} than the theoretical prediction near the peak, but is
consistently {\em smaller} than the theoretical prediction in the tails.
Central values of the QCD parameters have been used.\label{sum_rules_single}}
\end{figure}

Due to the instability of the $k=-1$ Gaussian sum-rule analysis, we turned to 
the $k=0$ sum-rule. We showed that hadronic parameters extracted using 
a single narrow resonance model led to poor agreement between
theory and phenomenology as indicated by the plots of Figures~\ref{sigma} 
and~\ref{sum_rules_single}.
Consequently, we then considered phenomenological models which allowed for 
resonance strength to be distributed over an appreciable energy range.

\begin{figure}[htb]
\centering
\includegraphics[scale=0.65]{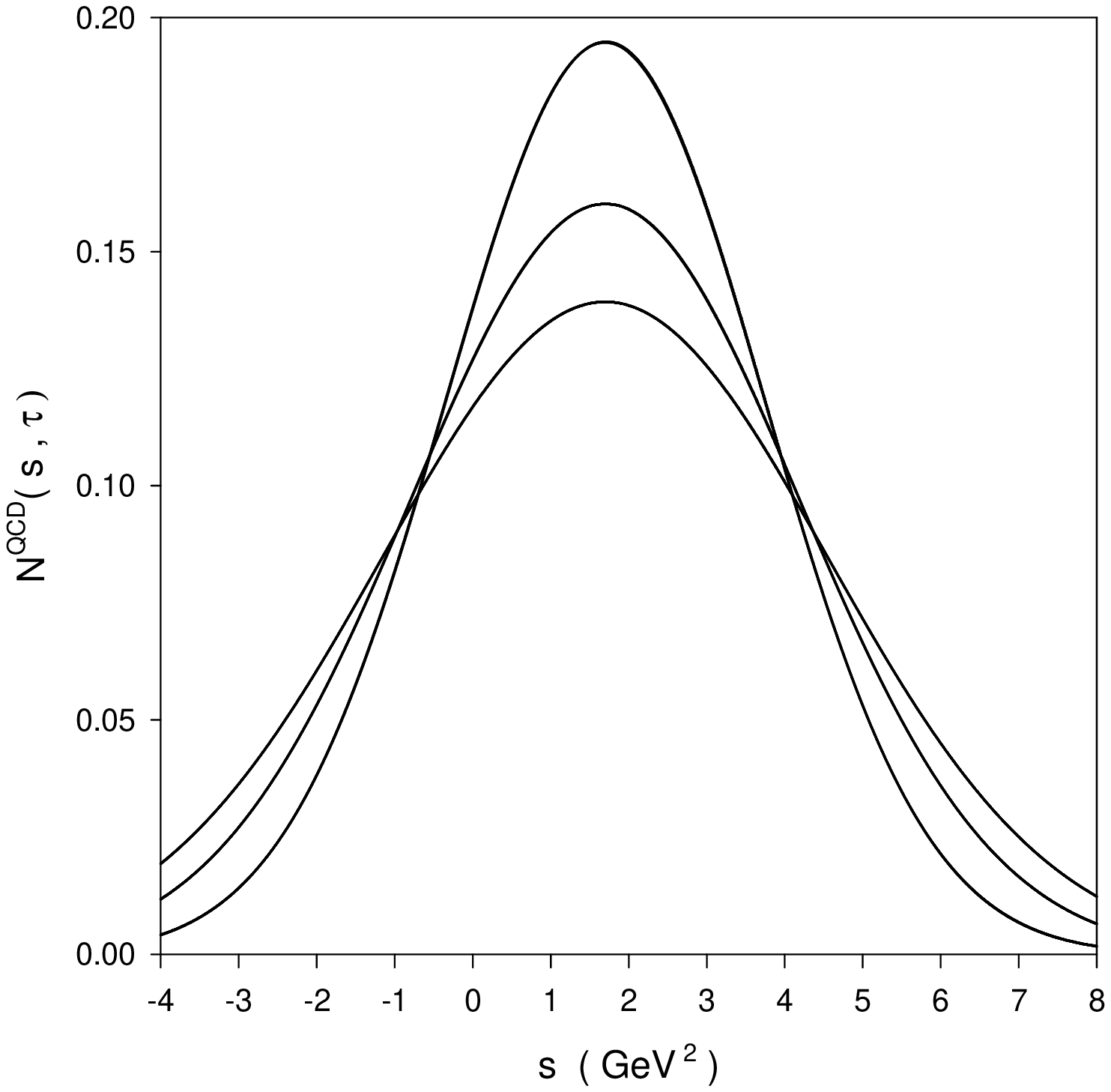}
\caption{
Typical comparison of the
theoretical prediction  $N_0^{QCD}\left(\hat s, \tau,s_0\right)$ with the
phenomenological side of the normalized Gaussian sum-rule (\protect\ref{ngsr})
in the single non-zero width resonance models and the double resonance models. 
Individual plots for each model are indistinguishable from those shown, and
are hence not repeated as additional Figures.
The $\chi^2$-optimized value of the
 continuum ($s_0=2.3\,{\rm GeV^2}$) has been used to extract the resonance
 parameters as
outlined in the text.
The $\tau$ values used for the three pairs of curves, from top to bottom in the figure, are respectively
 $\tau=2.0\ \unit{GeV}^4$, $\tau=3.0\ \unit{GeV}^4$, and $\tau=4.0\ \unit{GeV}^4$.
Note the almost perfect overlap between the theoretical prediction and the 
phenomenological models.
Central values of the QCD parameters have been used.}
\label{square_fig}
\end{figure}

We focused on three single non-zero width resonance models
(see Section~\ref{analysis}), and
the results of all the corresponding analyses were indicative of a 
 ($m\approx1.2\mbox{--}1.3 $ GeV), wide ($\Gamma\approx0.4\mbox{--}0.6$ GeV) 
resonance (see Table~\ref{width_tab}). 
The coincidence between plots of the theoretical and phenomenological Gaussian 
sum-rules (characterized by Figure~\ref{square_fig}) represented a vast improvement 
over the corresponding graphs obtained from a single narrow resonance analysis
(Figure~\ref{sum_rules_single}).
However, as indicated in Table~\ref{width_tab}, all three models failed 
completely to predict the value of the first independent moment 
combination $A_0^{(3)}$, prompting us to move on to 
analyses of phenomenological models that allow for a second resonance.

Regarding both a $\chi^2$ measure of the discrepancy between theoretical and 
phenomenological $k=0$ Gaussian sum-rules and the ability to accurately 
predict higher order, independent moments, the double resonance models 
represented a marked improvement compared to the single non-zero width 
resonance models. 
Resulting hadronic parameter estimates from the various models analyzed 
are summarized in Table~\ref{res_summary_table}.
Estimates of the larger of the two masses ($m$) were remarkably insensitive to 
the particular width model used, approximately 1.4 GeV in all cases considered. 
Mass predictions for the lighter of the two resonances ($M$) 
ranged from 1.0 GeV in the double narrow and narrow plus skewed Gaussian models  
to roughly 1.25 GeV in the narrow plus square and narrow plus Gaussian 
models---again, a surprising degree of stability considering the variety
of width models employed.
Further, in those models which allowed for a non-zero resonance width, 
we made no \textit{a priori} assumptions as to which of the two resonances was
wide; however, the width was consistently found to be associated with the lighter of the two states
and was relatively large in magnitude, yielding 
an equivalent Breit-Wigner width of $\Gamma_{\text{BW}}\approx0.55\,{\rm GeV}$
in the Gaussian resonance models.
These hadronic parameter estimates were subsequently employed in predictions 
of higher order, independent moment combinations. The results of this analysis
are summarized in Table~\ref{moms_summary_table}.
The deviation of the phenomenological predictions from the values computed
using QCD was typically  20--50\%, with the best agreement occurring in the
skewed Gaussian resonance model. While certainly not perfect,
this represented a vast improvement over analogous calculations
performed in the various single non-zero width resonance 
models (see Table~\ref{width_tab}).
Finally, in Table~\ref{chi_table}, we have collected the results of a
$\chi^2$ measure of the difference between the theoretical $k=0$ Gaussian sum-rules
and the various phenomenological sum-rules corresponding to each of the resonance
models considered. For the double resonance models, the values quoted 
were consistently an order of magnitude lower than those pertaining to the 
single non-zero width models.
Therefore, in addition to our previous conclusion concerning the necessity for 
distributed resonance strength, we note further that a single non-zero width resonance
model is insufficient to account for this spreading and that far better
consistency between theory and phenomenology (as encapsulated in the Gaussian sum-rules
of scalar gluonium) is achieved by employing models which allow for a second 
resonance.

Although certainly relevant to the nature of the spectrum of scalar states with  
masses under 2 GeV, it is difficult to make a direct comparison between our results 
and the entries of the PDG~\cite{pdb} since the typical mass separation between the various observed scalar
states is relatively small: on the order of 0.2 GeV.
Unfortunately, this is approximately the same magnitude as the uncertainties we calculate in our
mass estimates. 
These error bars are the direct result of uncertainties 
in the QCD parameters~(\ref{dimfour},\ref{DILparams}).
Therefore, in light of these uncertainties, we refrain from drawing definitive 
conclusions concerning the nature of any specific scalar state.  
Instead, we merely note that certain classification schemes proposed by 
Minkowski and Ochs~\cite{f0_1370_interp} and Narison~\cite{narison98}
require a relatively light ($\approx$ 1 GeV), wide state---a situation
completely consistent with our results. Furthermore, the persistent 
$m=1.4$ state we find may indeed have ramifications regarding the nature  
of either the $f_0(1370)$ or the $f_0(1500)$~\cite{f0_1500_interp}.

\smallskip
\noindent
{\bf Acknowledgements:}  The authors  are grateful for research support
from the Natural Sciences and Engineering Research Council of Canada (NSERC).

\clearpage

\clearpage

\begin{thebibliography}{99}

\bibitem{NSVZ_glue}
V.A.\ Novikov, M.A.\ Shifman, A.I.\ Vainshtein and V.I.\ Zakharov,
      Nucl.\ Phys.\ \tref{B165}{1980}{67}.

\bibitem{sum_rules_for_scalar_glue}
M.A. Shifman, Z. Phys.\ \tref{C9}{1981}{347};\\
P. Pascual and R. Tarrach, Phys.\ Lett.\ \tref{B113}{1982}{495};\\
S. Narison, Z. Phys.\ \tref{C26}{1984}{209};\\
C.A.\ Dominguez and N.\ Paver, Z.\ Phys.\ \tref{C31}{1986}{591};\\
J.\ Bordes, V.\ Gim\`{e}nez and J.A.\ Pe\~{n}arrocha, Phys.\ Lett.\ \tref{B223}{1989}{251};\\
J.L.\ Liu and D.\ Liu, J.\ Phys.\ \tref{G19}{1993}{373};\\
L.S.\ Kisslinger, J.\ Gardner and C.\ Vanderstraeten, Phys.\ Lett.\ \tref{B410}{1997}{1};\\
Tao Huang, Hong Ying Jin and Ai-lin Zhang, Phys.\ Rev.\ \tref{D59}{1998}{034026}.

\bibitem{two_res_glue}
S. Narison and G. Veneziano, Int.\ J.\ Mod.\ Phys.\ \tref{A11}{1989}{2751};\\
E.\ Bagan and T.G.\ Steele, Phys.\ Lett.\ \tref{B243}{1990}{413}.

\bibitem{narison98}
S.\ Narison, Nucl.\ Phys.\ \tref{B509}{1998}{312}.

\bibitem{shuryak_forkel}
E.V. Shuryak, Nucl.\ Phys.\ \tref{B203}{1982}{116};\\
Hilmar Forkel, hep-ph/0005004;\\
Hilmar Forkel, hep-ph/0103204.

\bibitem{har00}
D. Harnett, T.G. Steele and V. Elias, Nucl.~Phys.\ \tref{A686}{393}{2001}.

\bibitem{basic_instanton}
A.\ Belavin, A.\ Polyakov, A.\ Schwartz and Y.\ Tyupkin,
   Phys.\ Lett.\ \tref{B59}{1975}{85};\\
G.\ 't Hooft, Phys.\ Rev.\ \tref{D14}{1976}{3432};\\
C.G.\ Callan, R.\ Dashen and D.\ Gross, Phys.\ Rev.\ \tref{D17}{1978}{2717};\\
M.A. Shifman, A.I. Vainshtein and V.I. Zakharov, Nucl.~Phys.\ \tref{B165}{1980}{45}.

\bibitem{gauss}
R.A. Bertlmann, G. Launer and E. de Rafael, Nucl.~Phys.~\tref{B250}{1985}{61}.

\bibitem{orl00}
G. Orlandini, T.G. Steele and D. Harnett, Nucl.~Phys.\ \tref{A686}{261}{2001}.

\bibitem{let}
V.A. Novikov, M.A. Shifman, A.I. Vainshtein and V.I. Zakharov, Nucl.~ Phys. \tref{B191}{1981}{301}.

\bibitem{pdb}
D.E. Groom {\it et al}, Eur.~Phys.~J.\ \tref{C15}{2000}{1}.

\bibitem{che97}
K.G. Chetyrkin, B.A. Kneihl and M. Steinhauser, Phys.~Rev.~Lett.\ \tref{79}{353}{1997}.

\bibitem{bag90}
E. Bagan and T.G. Steele, Phys.~Lett.\ \tref{B234}{1990}{135}.

\bibitem{schaefer_shuryak}
T. Sch\"{a}efer and E.V. Shuryak, Phys.~Rev.~Lett.\ \tref{75}{1995}{1707}.

\bibitem{inst_K2}
B.V. Geshkenbein and B.L. Ioffe, Nucl.~Phys.\ \tref{B166}{1980}{340};\\
B.L. Ioffe and A.V. Samsonov, Phys.~of~Atom.~Nucl.\ \tref{63}{2000}{1527}.  

\bibitem{abr}
M. Abramowitz and I.E. Stegun, {\em Mathematical Functions with Formulas,
   Graphs, and Mathematical Tables} (National Bureau of Standards
   Applied Mathematics Series, Washington) 1972.

\bibitem{nar81}
S. Narison and E. de Rafael, Phys.~Lett. \tref{B103}{1981}{57}.

\bibitem{ste98}
T.G. Steele and V. Elias, Mod.~Phys.~Lett.\ \tref{A13}{1998}{3151}.

\bibitem{che98}
K.G. Chetyrkin, B.A. Kneihl and M. Steinhauser, Nucl.~Phys.\ \tref{B510}{1998}{61}.

\bibitem{fesr}
R.\ Shankar, Phys.\ Rev.\ \tref{D15}{1977}{755};\\
R.G.\ Moorhouse, M.R.\ Pennington, G.G.\ Ross, Nucl.\ Phys.\ \tref{B214}{1977}{285};\\
K.G.\ Chetyrkin, N.V.\ Krasnikov, N.N.\ Tavkhelidze Phys.\ Lett.\ \tref{76B}{1978}{83};\\
E.G.\ Floratos, S.\ Narison, E.\ de Rafael, Nucl.\ Phys.\ \tref{B155}{1979}{115}.

\bibitem{nar97}
S. Narison, Nucl.~Phys.~B~(Proc.~Supp.) \tref{54A}{1997}{238}.

\bibitem{SR}
M.A. Shifman, A.I. Vainshtein and  V.I. Zakharov, Nucl.~Phys.\ \tref{B147}{1979}{385,448};\\
L.J. Reinders, H. Rubenstein and S. Yazaki, Phys.~Rep.\ \tref{1}{1985}{1}.

\bibitem{bag85}
E. Bagan, J.I. Latorre, P.Pascual and T. Tarrach, Nucl.~Phys.\ \tref{B254}{1985}{555}.

\bibitem{DIL}
E.V. Shuryak, Nucl.~Phys.\ \tref{B203}{1982}{93}.


\bibitem{Elias}  
 V.\ Elias, A.H.\ Fariborz, M.A.\ Samuel, Fang~Shi, T.G.\ Steele,
Phys.\ Lett.\ \tref{B412}{1997}{131};\\
V.\ Elias, A.H.\ Fariborz, Fang~Shi, T.G.\ Steele, Nucl.\ Phys.\ \tref{A633}{1998}{279}.


\bibitem{pion} V.A. Novikov, M.A. Shifman, Z. Phys. \tref{C8}{1981}{43};\\
M.A. Shifman, Z. Phys. \tref{C9}{1981}{347}.

\bibitem{f0_1370_interp}
P. Minkowski and W. Ochs, Eur.~Phys.~J \tref{C9}{1999}{283}.


\bibitem{f0_1500_interp}
C. Amsler and F.E. Close, Phys.~Rev.\ \tref{D53}{1996}{295};\\
L. Burakovsky and P.R. Page,  Phys.~Rev.\ \tref{D59}{1999}{014022}.

\end{thebibliography}
\end{document}